\pdfoutput=1
\documentclass[IEEEtran,print]{ieeecolor}

\usepackage{jsen}

\usepackage{cite}
\usepackage{amsmath,amssymb,amsfonts}
\usepackage{algorithmic}
\usepackage[ruled,vlined,linesnumbered]{algorithm2e}
\usepackage{graphicx}
\usepackage[rgb]{xcolor}
\usepackage{url}
\usepackage{wrapfig}
\usepackage{pifont}
\usepackage{booktabs}
\usepackage{footnote}
\usepackage{mathtools}
\usepackage{stmaryrd}
\usepackage{breqn}
\usepackage{dsfont}

\SetAlCapNameFnt{\footnotesize}
\SetAlCapFnt{\footnotesize}
\SetAlgorithmName{Algorithm}{Algorithm}{List of Algorithms}

\usepackage{multirow}
\usepackage{tabularx}
\usepackage{adjustbox}
\usepackage{rotating}
\usepackage{stfloats}
\usepackage{float}
\usepackage{subfig}
\usepackage{makecell}
\usepackage{boldline}

\usepackage{titlesec}
\usepackage[hang,flushmargin]{footmisc}

\usepackage{enumerate}

\usepackage{nomencl}
\makenomenclature

\newcommand{\cmark}{\ding{51}} 
\newcommand{\xmark}{\ding{55}} 

\makeatletter
\newcommand{\algorithmfootnote}[2][\footnotesize]{%
  \let\old@algocf@finish\@algocf@finish
  \def\@algocf@finish{\old@algocf@finish
    \leavevmode\rlap{\begin{minipage}{\linewidth}
    #1#2
    \end{minipage}}%
  }%
}
\makeatother

\usepackage{subfig}
\usepackage{titlesec}
\usepackage{multicol}

\newcommand{\unc}[1]{\,{\scriptstyle \pm}\,#1}
\usepackage{tabularx,ragged2e}
\newcolumntype{L}{>{\RaggedRight\arraybackslash}X}
\usepackage{comment}

\def\BibTeX{{\rm B\kern-.05em{\sc i\kern-.025em b}\kern-.08em
    T\kern-.1667em\lower.7ex\hbox{E}\kern-.125emX}}

\IEEEtitleabstractindextext{%
\begin{abstract}
\begin{wrapfigure}{r}{0.6\textwidth}
    \vspace{-5pt}
    \centering
    \includegraphics[width=0.6\textwidth]{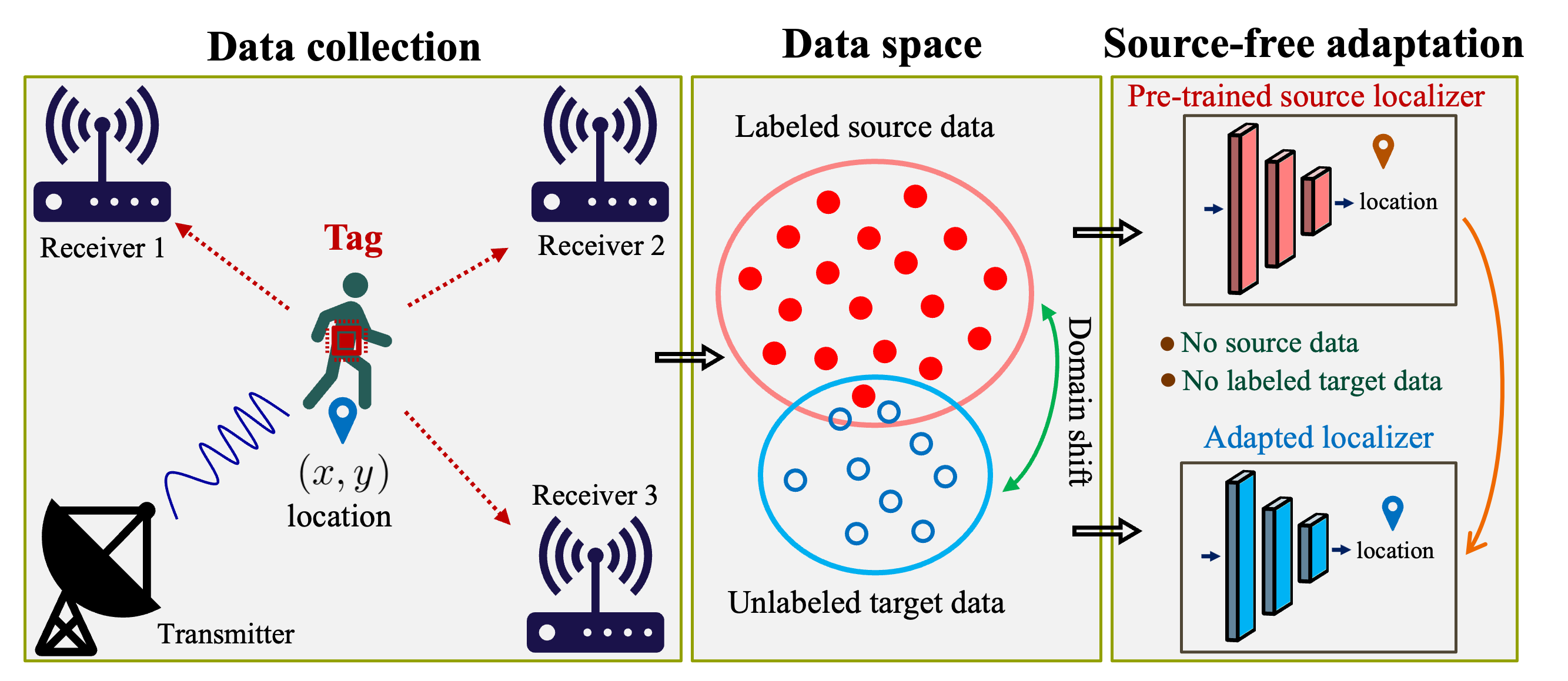}
    \vspace{-18pt}
\end{wrapfigure}

Indoor localization using radio-frequency identification (RFID) has benefited from deep learning, yet models trained in a labeled source environment often degrade when deployed in a different, unlabeled target environment. Unsupervised domain adaptation (UDA) aims to mitigate this distribution shift by aligning a source-trained model with target-domain data. In practice, the source dataset is frequently unavailable at adaptation time due to privacy and resource constraints. This motivates source-free domain adaptation (SFDA); however, most SFDA methods have been developed for classification, and extending them to indoor localization (regression) is challenging, especially when target datasets are small and noisy. Motivated by above limitations, we introduce MTLoc, a source-free mean-teacher approach for indoor localization. MTLoc includes a student and a teacher network: the student network is updated using noisy target data with teacher-generated pseudo-labels. The teacher network maintains stability through exponential moving averages. To further ensure robustness, we propose a correction mechanism in which the teacher's pseudo-labels are refined using $k$-nearest neighbor correction. MTLoc allows for self-supervised learning on target data, facilitating effective adaptation to dynamic and noisy indoor environments. Validated using real-world data from our experimental setup with INLAN Inc., our results\textsuperscript{1} show that MTLoc achieves high localization accuracy under challenging conditions, significantly reducing distance error compared to baselines. On average, it reduces MAE by $20.0\%$ on Cross and $22.5\%$ on Square datasets. With confidence correction, these improvements reach $23.9\%$  and $28.2\%$  respectively.

\end{abstract}

}

\title{MTLoc: A Confidence-Based Source-Free Domain Adaptation Approach For Indoor Localization}

\author{
\begin{minipage}{\textwidth}
\centering
Negar Mehregan, Berk Bozkurt, Eric Granger,\\
Mohammadjavad Hajikhani, and Mohammadhadi Shateri
\end{minipage}
\thanks{N. Mehregan, E. Granger, and M. Shateri are with the LIVIA, Department of Systems Engineering, ETS Montreal, Canada (e-mails: negar.mehregan.1@ens.etsmtl.ca, eric.granger@etsmtl.ca, mohammadhadi.shateri@etsmtl.ca).}%
\thanks{B. Bozkurt and M. Hajikhani are with INLAN Inc., Montreal, Canada (e-mails: b.bozkurt@inlantech.com, m.hajikhani@inlantech.com).}%
}

\begin{document}

\maketitle
\footnotetext[1]{Code and datasets: \url{https://github.com/negarmehregan/Indoor-Localization/tree/main}}

\IEEEpeerreviewmaketitle

\includegraphics[width=0.2\textwidth]{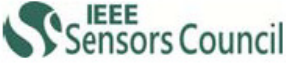}

{\footnotesize

\begin{tabularx}{\columnwidth}{@{} l L @{}}
\multicolumn{2}{c}{\MakeUppercase{Nomenclature}}\\[4pt]
 $X, Y$ & Random variables \\
 $x, y$ & Realizations for $X, Y$\\
 $p(x)$ & Probability distribution of $X$ \\
 $p(x \mid y)$ & Conditional distribution of $X$ given $Y$ \\
 $\mathbb{E}[X]$ & Expectation of the random variable $X$ \\
 EMA ($\alpha$) & Exponential moving average with parameter $\alpha$ \\
 GRL & Gradient Reversal Layer \\
 $k$-NN & $k$-nearest-neighbor \\
 
\end{tabularx}
}

\section{Introduction}

\IEEEPARstart{R}{obust} and reliable localization systems aim to revolutionize positioning in fields like the Internet of Things \cite{10416167}, and Augmented Reality \cite{10269042}. Various indoor localization techniques are being studied, in which WiFi-based methods are attracting the most attention due to their compatibility, cost-effectiveness, and wide applicability \cite{8879523,8918264,9613334}. 

As GPS signals are proven to be ineffective for localization due to signal deflection and multipath-related issues \cite{10706359}, different technologies are introduced, with qualifications varying based on the type of tag applied. Notably, ultra-wideband (UWB) tags \cite{10756560,huang2023indoor} with high-precision accuracy and low power usage, radio frequency identification (RFID) tags \cite{7842591,yang2014tagoram} with lower cost and long-range communication, and Bluetooth low energy (BLE) tags \cite{9075151,9343321}, a medium range, energy-efficient technology, are used in localization models. 

Tag-based tracking models are used to monitor key assets' location, ensure safety, and localize objects for healthcare applications \cite{6587041,5162266,5304900}. A typical indoor positioning system includes receivers, transmitters, and tags for data collection, where a signal is transmitted to the receiver after interacting with the tagged object. In various studies, through using receivers, different types of data were acquired from tags including received signal strength \cite{10000400,10190065,10705937}, Phase, Channel Impulse Response \cite{9903809,10006709}, channel state information (CSI) \cite{Rao2024,10838284}, angle of arrival \cite{10472153,10260273,9676583}, and radio frequency \cite{9244574,9555623}. However, the inaccuracies caused by signal attenuation and distortion remain a challenge, particularly in complex environments in which multipath components and non-line-of-sight effects stand out as the most concerning \cite{Wang2024,10006709}. To address these issues, various deep learning-based techniques are proposed for indoor localization.

Previous studies highlight the impact of environmental change on data \cite{chen2020FiDO,li2021DAFI}. As mentioned by \cite{Yan2024}, CSI data is affected by any changes in the environment, hence, a localization system might not perform well in a new environment. Therefore, unsupervised domain adaptation (UDA) methods have been proposed, where the source domain model is fine-tuned to be applicable in a distinct but related target domain without access to ground truth. Given a pre-trained model utilizing labeled source data collected from Warehouse A, the objective is to evaluate its performance in a different environment, Warehouse B. The primary challenge lies in the fact that the new domain (Warehouse B) may produce substantially different object-related signals, possibly due to variations in the positions of transmitters and receivers. Retraining the model is infeasible due to the lack of annotated data from Warehouse B, falling under the framework for UDA.

Nevertheless, UDA raises concerns regarding data privacy, particularly when sensitive information is present in the source domain \cite{9200758}. To avoid data leakage, source-free domain adaptation (SFDA) has been proposed, which enables robust adaptation without access to source data. SFDA not only respects privacy concerns but also reduces computational costs and facilitates storage and transfer \cite{Fang2023,10452835}. Additionally, regulatory constraints or corporate policies often restrict data transfer, further motivating SFDA as a practical approach \cite{10452835}. The scalability of SFDA also supports adaptation across diverse environments \cite{10452835,10078842}.
While most SFDA methods target classification, TASFAR \cite{10597720} is a recent regression-based approach that estimates the target label distribution by modeling predictive uncertainty via a label-density map. It generates pseudo-labels for target samples to finetune the source model.

To the best of our knowledge, no method for SFDA has been proposed for indoor localization. Of the current SFDA methods, the mean teacher framework is particularly well-suited for regression-based indoor localization. In practice, target datasets for indoor localization are often limited and noisy, making domain adaptation reliant on highly confident pseudo-labels. Researchers such as Yeonguk et al. \cite{Yu2024} have used this technique due to its EMA-based stabilization, which produces stable pseudo labels through consistency regularization. Although it is a simple approach, it ensures efficient performance in source-free scenarios while leveraging self-supervision. 

Motivated by these characteristics, we introduce mean teacher localization (MTLoc) for SFDA, a self-supervised method leveraging a mean teacher framework. This approach uses a teacher-student architecture, where both networks are initialized with a localizer pre-trained on source data. The student model is fine-tuned on noisy target domain data using pseudo labels generated by the teacher network, enabling effective adaptation without access to source data. To further improve performance, we propose an enhanced model, MTLoc with confidence, where the teacher's uncertain pseudo-labels are corrected using $k$-nearest neighbors ($k$-NN), and the student model is trained exclusively on the most confident ones. This refinement improves reliability by focusing on high-certainty predictions, reducing variations, and enhancing robustness. 

The \textbf{main contributions} of this paper are as follows.
\begin{itemize}
   \item We present the MTLoc, an SFDA method for indoor localization, by leveraging a self-supervised knowledge distillation framework to adapt from small and noisy target data.   
    \item An improved confidence-based variant of MTLoc is proposed, where uncertain pseudo labels generated by the teacher network are refined using a $k$-NN strategy to compute a weighted average of the $k$ most confident pseudo labels. This ensures that the student network is trained exclusively on high-confidence labels, enhancing overall model robustness.
    \item A comprehensive data collection process was conducted to develop three datasets—Ceiling (source), Cross, and Square (targets)—at the INLAN Inc. testing facility. This setup involved four receivers, a transmitter equipped with dipole antennas, and an RFID tag, providing a robust evaluation environment for the proposed method. MTLoc is validated on this real-world data and significantly outperforms the state-of-the-art UDA and SFDA models.
\end{itemize}

This paper is organized as follows. Sections \ref{sec:sectionII_new} provides related work. Section \ref{sec:sectionII} defines the SFDA problem and baseline. Section \ref{sec:sectionIII} presents our proposed confidence-based SFDA method. Section \ref{sec:sectionIV} details data collection, experimental protocol, performance measures, and results analysis.

\section{Related Work}\label{sec:sectionII_new}

Over the years, several approaches have been proposed for indoor localization. Some rely on kernel-based methods, such as Zhang et al.~\cite{Zhang2021IMKELM}, who developed an integrated multiple-kernel extreme learning machine that uses temporal and spatial CSI features for device-free localization in cluttered environments. However, AI-based techniques have become the dominant trend. 
For instance, with MFFALoc~\cite{Rao2024}, CSI fingerprints are fed into a convolutional neural network (CNN), tackling multipath and non-line-of-sight effects. Rao et al.~\cite{Rao2024} further remove environmental noise by utilizing waveform diversity and detection, and implement a feature fusion system to merge multi-antenna feature matrices, resulting in fine-grained fingerprints. For example, MFFALoc~\cite{Rao2024} use a CNN on CSI fingerprints to tackle multipath and non-line-of-sight effects. In addition, Rao et al.~\cite{Rao2024} improve fingerprint quality by using waveform diversity and detection to suppress environmental noise, and by fusing multi-antenna feature matrices to construct fine-grained representations. In parallel, Zhang et al.~\cite{10930823} address domain adaptation by using online learning, where fingerprint augmentation and a sequential extreme learning machine with a forgetting mechanism allow the model to adapt dynamically to environmental changes.
Some studies, such as Chen et al. \cite{9082193}, apply 1D-CNN for RF-based environment classification, while others, like Yan et al. \cite{9541367}, use 2D deep CNNs for CSI feature extraction to improve localization accuracy. Se-Loc~\cite{Ye2023} is another CNN-driven approach with a security-enhanced semi-supervised indoor localization technique, integrating a denoising autoencoder (AE) to improve fingerprinting. This model employs a Pearson correlation coefficient-based access point selection system to eliminate contaminated access points. Cao et al. \cite{113645} proposed a confidence-aware mean teacher framework for metallographic image semantic segmentation, which adaptively calibrates pseudo-labels with class-aware thresholds and improves data augmentation through confidence-driven patch mixing.

Some research works, such as~\cite{Ruan2023}, use AE for spatial feature extraction. Likewise, in~\cite{Li2023} and~\cite{Wang2024}, a variational AE is applied to extract key position-related features and reduce environmental noise to solve localization and multipath challenges. Some works merge recurrent neural networks with AE, such as ~\cite{Ayinla2024}, where a stacked AE is combined with an attention-based long short-term memory network, enhancing localization accuracy. Liu et al.\cite{Liu2024} use gated recurrent units for feature extraction, while Chen et al.\cite{8733822} apply a deep LSTM with a local feature extractor to process RSSI fingerprints, capturing temporal dependencies through a regression layer. Generative models have also been explored for data augmentation~\cite{Junoh2024, Li2021, Wu2023}, improving localization accuracy. 

In fault diagnosis settings, Wang et al.~\cite{wang2025domain} provide a comprehensive survey of domain adaptation,  focusing on both homogeneous methods, including metric learning, adversarial learning, and feature reconstruction, along with heterogeneous settings, source-free, domain generalization, partial, open-set, and universal domain adaptation methods. 

Currently, UDA approaches can be classified into three main categories \cite{Li2024}: (a) domain distribution alignment \cite{Long2015,Cui2020,Shui2023}, (b) adversarial learning to extract domain-invariant representation \cite{Long2018,Tzeng2017,Li2022}, and (c) model fine-tuning through self-supervision \cite{Lian2019,Zou2018,Li2022Divergence}. In indoor localization, most UDA approaches rely on adversarial learning through gradient reversal layers (GRL). Examples include Zhang et al.~\cite{Zhang2023}, who use a semi-supervised graph convolutional network with RSSI data; Xu et al.~\cite{Xu2024}, who propose a Gaussian noise–infused co-teaching mechanism; Zhang et al.~\cite{10745164}, who develop dynamic adversarial adaptation networks to eliminate inconsistencies in global and local discriminators; and Prasad et al.~\cite{Prasad2024}, who apply an AE for dimension reduction alongside a GRL to create domain-invariant features. 

Discrepancy-based models aligning source and target distributions have also been explored, such as~\cite{Zhang2024a}, which uses the Grassmann manifold for dimensionality reduction while preserving data integrity, combined with maximum mean discrepancy for accuracy. Hybrid mechanisms combining generative and GRL-based approaches also exist, such as~\cite{Chen2022} and~\cite{Yan2024}, where generative models (variational AE and adversarial AE, respectively) are used for augmentation. In~\cite{Chen2022}, a domain-adaptive classifier includes a feature extractor with classification-reconstruction layers, while in~\cite{Yan2024}, a CNN feature extractor is coupled with GRL-based discriminator and predictor networks.

As discussed earlier, SFDA overcomes the limitations of UDA regarding privacy, storage, and transfer constraints. Fang et al.~\cite{Fang2023} provide a detailed survey of SFDA in image-based studies, highlighting its efficiency and scalability. However, no SFDA methods have yet been applied to indoor localization, leaving an open gap that this paper addresses.

\section{Problem Definition and Baseline Overview}\label{sec:sectionII}

\subsection{Problem Formulation:}

Consider an indoor localization framework consisting of $N$ receivers and a single transmitter. The transmitter emits signals that interact with objects within the environment, and these signals are subsequently captured by the receivers. The objective is to develop an object localization system by analyzing the signals recorded by the receivers. Let $Z$ denote the random variable representing the signals recorded by the receivers, and let $(X, Y)$ represent the joint random variables corresponding to the true location associated with the recorded signal $Z$. The goal is to learn an object localizer defined as the mapping $\mathcal{M}(\cdot; \theta): \mathcal{Z} \rightarrow \mathcal{X} \times \mathcal{Y}$ parameterized by $\theta$, which produces an approximation $(\hat{X}, \hat{Y})$ of the true location for each observed signal $Z$, i.e $(\hat{X}, \hat{Y}) = \mathcal(Z; \theta)$. To optimize the parameters $\theta$ of the localizer, we employ Bayesian optimization by maximizing the log-likelihood of the true location given the estimated location as follows:
\begin{equation}\label{eq1}
\theta^{*} = \arg\max_{\theta} \log p_{X, Y \mid \hat{X}, \hat{Y}}(x, y \mid \hat{x}, \hat{y}) , 
\end{equation}
\noindent where depending on the assumed statistical model of the localization errors, this maximum likelihood estimation leads to different loss functions.

\noindent\textit{Gaussian Error Model}\cite{bishop2006pattern}:
Assuming the localization errors $(X - \hat{X}, Y - \hat{Y})$ follow a bivariate Gaussian distribution with zero mean and covariance matrix $\Sigma$, the negative log-likelihood function becomes:
\begin{equation}
- \log p(x, y \mid \hat{x}, \hat{y}) \propto \begin{pmatrix} x - \hat{x} \\ y - \hat{y} \end{pmatrix}^\top \Sigma^{-1} \begin{pmatrix} x - \hat{x} \\ y - \hat{y} \end{pmatrix}.
\end{equation}
If $\Sigma$ is proportional to the identity matrix, the expression simplifies, and maximizing the likelihood presented in Equ.~\eqref{eq1} is equivalent to the following problem:
\begin{equation} \label{eq:gaus}
\theta^{*} = \arg\min_{\theta} \mathbb{E}_{p_{X,Y,Z}}\left[ (X - \hat{X})^2 + (Y - \hat{Y})^2 \right] , 
\end{equation}
\noindent where the expectation is taken over the joint distribution of $p_{X,Y,Z}$.

\noindent\textit{Laplacian Error Model}\cite{bishop2006pattern}: Alternatively, considering the localization errors follow a bivariate Laplacian distribution and assuming independence between error in $X$ and $Y$, the joint probability density function can be written as follows:
\begin{equation}
p(x, y \mid \hat{x}, \hat{y}) = \left( \frac{1}{\sqrt{2}b} \right)^2 \exp\left( -\frac{| x - \hat{x} | + | y - \hat{y} |}{\sqrt{2}b} \right),
\end{equation}
\noindent where $b > 0$ is the scale parameter. In this case, the negative log-likelihood function is simplified to:
\begin{equation}
- \log p(x, y \mid \hat{x}, \hat{y}) = \frac{1}{\sqrt{2}b} \left( | x - \hat{x} | + | y - \hat{y} | \right ) + 2 \log (\sqrt{2}b).
\end{equation}
Ignoring the constant terms, maximizing the likelihood in Equ.~\eqref{eq1} corresponds to the following problem:
\begin{equation} \label{eq:lap}
\theta^{*} = \arg\min_{\theta} \mathbb{E}_{p_{X,Y,Z}}\left[ | X - \hat{X} | + | Y - \hat{Y} | \right].
\end{equation}

In this work, either formulation~\eqref{eq:gaus} or \eqref{eq:lap} could be used for optimizing the parameters $\theta$ of localizer $\mathcal{M}(\cdot; \theta)$, however, we applied equation \eqref{eq:lap}. After training $\mathcal{M}(\cdot; \theta)$ using data from a source domain $\mathbf{D}_s$ characterized by the joint distribution $p^s_{X,Y,Z}$, we aim to adapt it to a new target (operational) domain $\mathbf{D}_t$, where the joint distribution is $p^t_{X,Y,Z} \neq p^s_{X,Y,Z}$. The discrepancy between the source and target domains poses a challenge, as the localizer trained on $\mathbf{D}_s$ may perform poorly on $\mathbf{D}_t$ due to distribution shifts. Our objective is to adapt $\mathcal{M}(\cdot; \theta)$ to perform well in the target domain using an unlabeled $\mathbf{D}_t$.

\subsection{Baselines: }

\begin{figure}[!t] 
  \centering
  \includegraphics[width=1\linewidth]{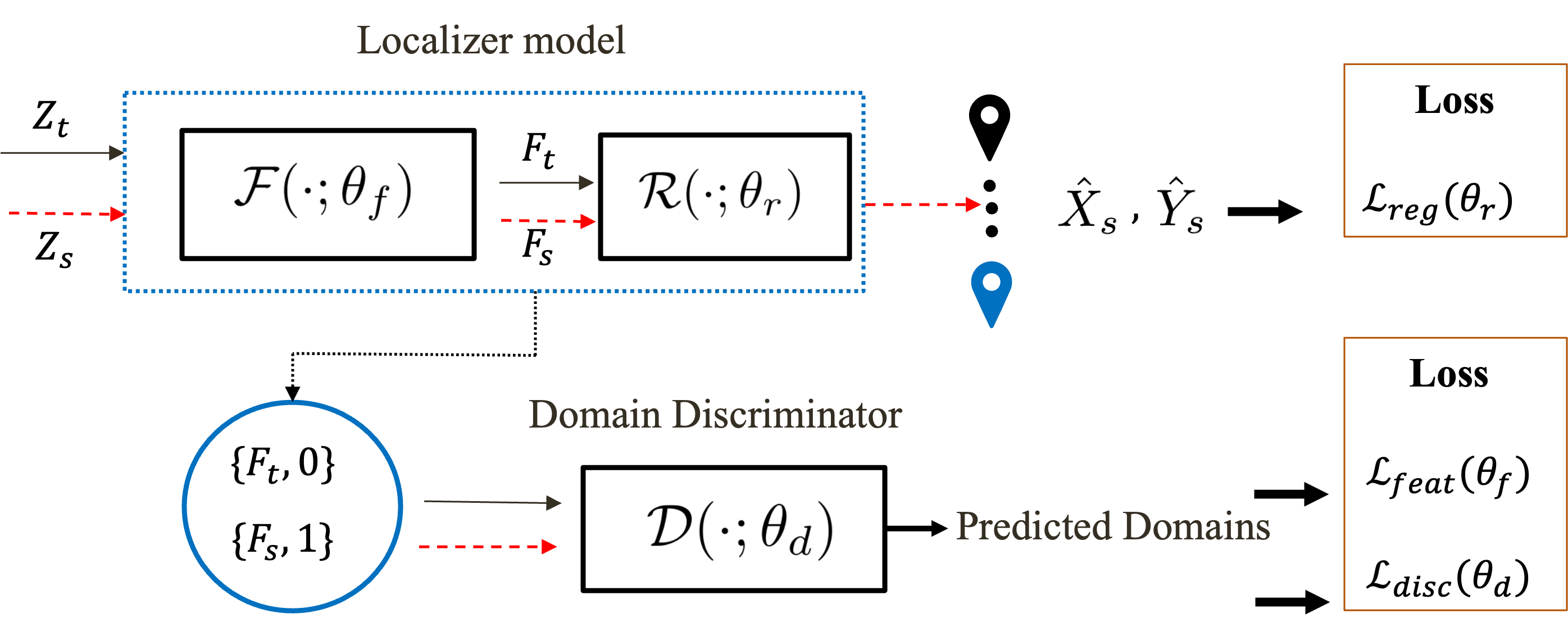} 
  \caption{\textit{System overview: }The source data \( Z_s \) and the unlabeled target data \( Z_t \) are processed through our Localizer model, which comprises a feature extractor \( \mathcal{F}(\cdot; \theta_f) \) and a regressor \( \mathcal{R}(\cdot; \theta_r) \). The source data \( Z_s \) is used to generate predictions, while the feature extractor \( \mathcal{F}(\cdot; \theta_f) \) is trained to deceive the discriminator \( \mathcal{D}(\cdot; \theta_d) \).}
\label{fig:DomainAdaptationBaseline}
\end{figure}

In this study, we considered several baselines including a source-only pre-adaptation model and an oracle approach trained on labeled target datasets \cite{varailhon2024sourcefreedomainadaptationyolo,Belal_2024_WACV}. Among UDA approaches created for indoor localization, adversarial UDA and discrepancy-based models are prominent, with the state-of-the-art focusing predominately on the use of GRL-based adversarial UDA techniques. Thereby, we evaluated an adversarial UDA method that leverages a GRL-based model. This approach is inspired by FreeLoc \cite{Yan2024}, an adaptation of Domain-Adversarial Neural Network (DANN) \cite{Ganin2015DANN}, which facilitates domain alignment through adversarial training, and serves as another baseline. As illustrated in Fig.~\ref{fig:DomainAdaptationBaseline}, the framework assumes access to a labeled source dataset, $\mathbf{D}_s=\{z_s^{(i)},(x_s^{(i)},y_s^{(i)})\}_{i=1}^{N_s}$ with distribution $p^s_{X,Y,Z}$, and an unlabeled target dataset, $\mathbf{D}_t=\{z_t^{(j)}\}_{j=1}^{N_t}$ with distribution $p^t_{Z}$. Here, $z$ represents the signals received by the receivers, while $(x, y)$ denotes the object location in the corresponding domain.

This framework leverages a localizer model that comprises two core components: a feature extractor $\mathcal{F}(\cdot; \theta_f)$, which maps input signals to a latent feature space, and a regressor $\mathcal{R}(\cdot; \theta_r)$, which predicts the object's location based on the extracted features. Additionally, a domain discriminator $\mathcal{D}(\cdot; \theta_d)$ is incorporated to facilitate domain adaptation. During the adaptation process, the feature extractor processes received signals from both the source and target domains, while the domain discriminator is trained to distinguish between features from these domains. Concurrently, the regressor is optimized using labeled source data to predict object locations accurately. The feature extractor, however, is adversarially trained to both confuse the domain discriminator and preserve localization accuracy for the regressor. 

The training objective is formalized such that the feature extractor aims to align the source and target domains in the feature space. This alignment occurs when the domain discriminator can no longer differentiate between the two domains, thus enabling the regressor trained on source data to generalize effectively to the target domain.
\begin{align} \label{eq:adversary}
\min_{\theta_f,\theta_r} \max_{\theta_d} \, & \mathbb{E} \left[ \log \mathcal{D}(\mathcal{F}(Z_s)) \right] + \mathbb{E} \left[ \log (1 - \mathcal{D}(\mathcal{F}(Z_t))) \right] \notag \\
& + \mathbb{E} \left[ | X_s - \hat{X}_s | + | Y_s - \hat{Y}_s | \right].
\end{align}
The first two terms represent the adversarial objective, where the feature extractor $\mathcal{F}$ seeks to mislead the domain discriminator $\mathcal{D}$. The final term represents the regression loss, encouraging the regressor $\mathcal{R}$ to minimize the distance between the predicted and true object locations for the source domain. Through this adversarial training, feature alignment between the source and target domains is achieved, enabling robust location prediction in the target domain. It should be mentioned that for training each model, the following loss functions are defined based on Equ.~\eqref{eq:adversary} where the expectations are approximated empirically. More precisely, let $\{z_s^{(b)},(x_s^{(b)},y_s^{(b)})\}_{b=1}^{B}$ and $\{z_t^{(b)}\}_{b=1}^{B}$ be a set of $B$ data samples from source and target respectively. The loss functions for training each network in Fig.~\ref{fig:DomainAdaptationBaseline} are presented in equations~\eqref{eq:regressor_loss}\eqref{eq:discriminator_loss}\eqref{eq:feature_loss}.
\begin{align} \label{eq:regressor_loss}
\mathcal{L}_{\text{reg}}(\theta_r) &= \mathbb{E}_{p^s_{X,Y,Z}} \left[ | X_s - \hat{X}_s | + | Y_s - \hat{Y}_s | \right]\notag\\
&\approx\frac{1}{B}\sum_{b=1}^{B} \left[ | x_s^{(b)} - \hat{x}_s^{(b)} | + | y_s^{(b)} - \hat{y}_s^{(b)} | \right],
\end{align}
\begin{align} \label{eq:discriminator_loss}
\mathcal{L}_{\text{disc}}(\theta_d) &= -\mathbb{E}_{p^s_{Z}} \left[ \log \mathcal{D}(\mathcal{F}(Z_s)) \right] - \mathbb{E}_{p^t_{Z}} \left[ \log (1 - \mathcal{D}(\mathcal{F}(Z_t))) \right]\notag\\
&\approx\frac{-1}{B}\sum_{b=1}^{B}\left[\log \mathcal{D}(\mathcal{F}(z_s^{(b)})) + \log (1 - \mathcal{D}(\mathcal{F}(z_t^{(b)})))\right],
\end{align}
\begin{align} \label{eq:feature_loss}
\mathcal{L}_{\text{feat}}(\theta_f) &= \mathcal{L}_{\text{reg}} - \mathcal{L}_{\text{disc}}.
\end{align}

The detailed training procedure of the feature extractor, regressor, and discriminator for our baseline model adversarial UDA is given in Algorithm~\ref{alg:baseline_alg1} which illustrates how these networks are updated depending on their functionality. In the next section, our proposed method for indoor localization will be discussed. Unlike state-of-the-art approaches, our UDA method is source-free, meaning that it does not rely on access to source data $\mathbf{D}_s$ for adaptation.

\begin{algorithm}
\small
\caption{Domain Adaptation with Access to Source Data} \label{alg:baseline_alg1}
\begin{algorithmic}[1]
\STATE \textbf{Input:} Labeled Source data$\ Z_s $, Target data $Z_t$
\STATE \textbf{Initialize:} Feature Extractor \( \mathcal{F}(\cdot; \theta_f) \), Regressor \( \mathcal{R}(\cdot; \theta_r) \), and Discriminator \( \mathcal{D}(\cdot; \theta_d) \)
\WHILE{not converged}
    \STATE Extract features: $F_s = F(Z_s; \theta_f)$, $F_t = F(Z_t; \theta_f)$
    
    \STATE Update $\theta_r$ by $ \nabla_{\theta_r}\mathcal{L}_{reg}$(Equ.~\eqref{eq:regressor_loss})
      
    \STATE Train Discriminator \( \mathcal{D}(\cdot; \theta_d) \) on $(F_s, F_t)$

    \STATE Update $\theta_F$ by $ \nabla_{\theta_f}\mathcal{L}_{feat}$ (Equ.~\eqref{eq:feature_loss})
    Received Signal Strength Indicator data 
    \STATE Update $\theta_D$  by $ \nabla_{\theta_d}\mathcal{L}_{disc}$ (equation ~\eqref{eq:discriminator_loss})
    
\ENDWHILE
\end{algorithmic}
\end{algorithm}

As additional UDA baselines, we consider several alignment techniques. ADDA \cite{Tzeng2017ADDA} performs adversarial marginal alignment by freezing the source encoder and training a target encoder adversarially against a domain discriminator so that target features match the source distribution, then reusing the source regressor. FiDO \cite{chen2020FiDO}, an autoencoder-inspired model, enforces reconstruction-based invariance by pairing a feature extractor and localizer with a parallel reconstruction head; it is trained on labeled source and unlabeled target data to learn domain-invariant features. DAFI \cite{li2021DAFI} applies class-conditional adversarial alignment with a shared feature extractor and localizer, using two GRL-trained domain discriminators (one feature-level and one class-conditioned).
In addition, we adapted SHOT \cite{pmlr-v119-liang20a}, an SFDA method originally designed for classification, for the regression task.

\section{The Proposed MTLoc Method}\label{sec:sectionIII}

\begin{figure}[htbp!] 
  \centering
  \includegraphics[width=1\linewidth]{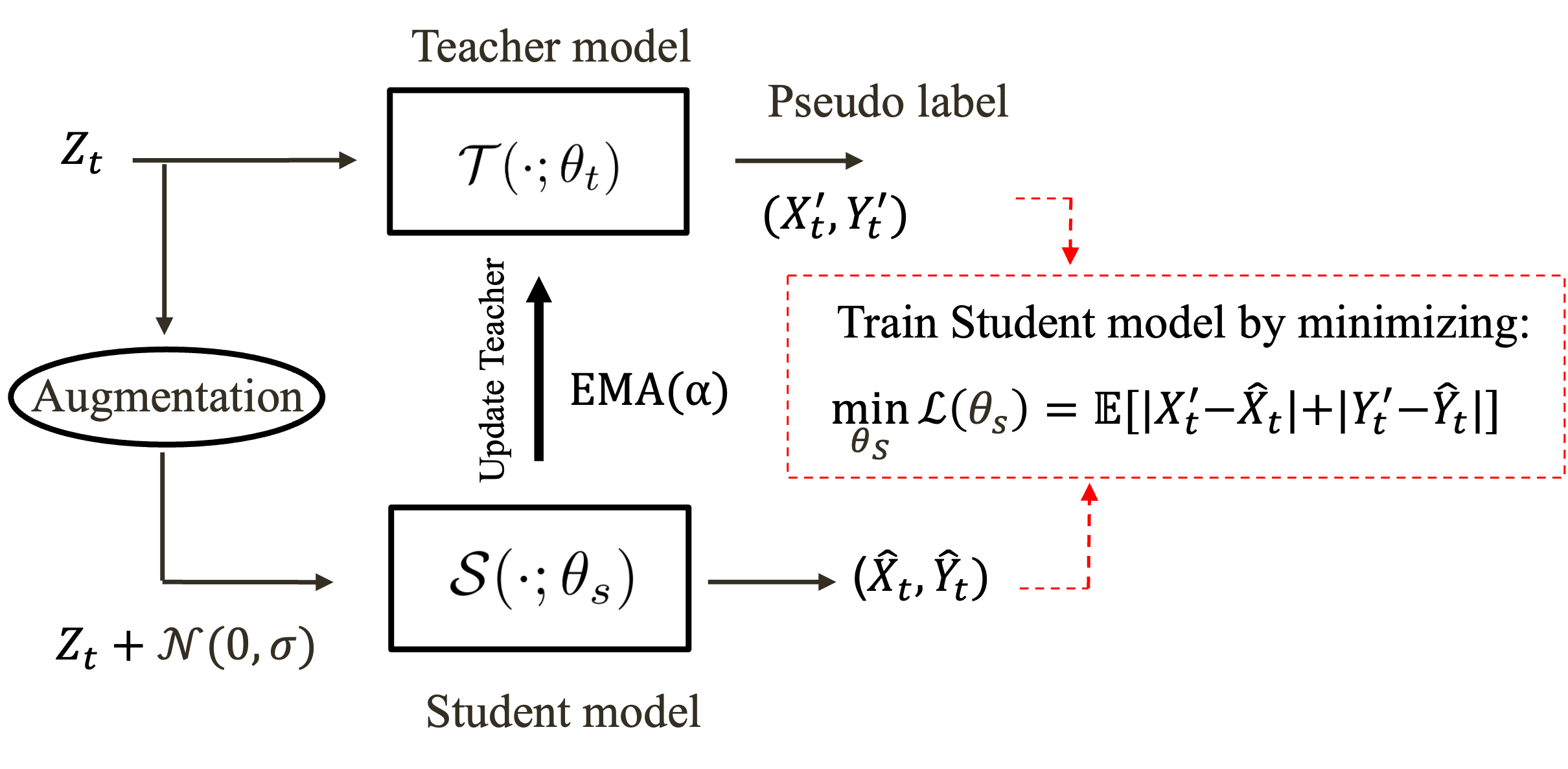} 
  \caption{\textit{MTLoc: The unlabeled target data ($Z_t$) is passed to the Teacher model \( \mathcal{T}(\cdot; \theta_t) \) to generate pseudo labels ($X_t',Y_t'$), while undergoing an augmentation process, through Gaussian Noise, allowing the Student model \( \mathcal{S}(\cdot; \theta_s) \) to proceed with the prediction $\hat{X_t},\hat{Y_t}$ while minimizing the Knowledge Distillation Loss $L_{\text{KD}}$.}}
\label{fig:MTLoc}
\end{figure}

In this study, an SFDA approach is proposed for indoor localization, called mean teacher localization (MTLoc), illustrated in Fig. \ref{fig:MTLoc}. As the current approaches are conditioned on the availability of the source domain data, making them not feasible in real-world scenarios~\cite{Fang2023}, we developed a source-free model. In this technique, two models are trained to ensure the quality and effectiveness of the training. This approach implements two networks: 1) teacher network \( \mathcal{T}(\cdot; \theta_t) \)~\cite{Fang2023} 2) student network \( \mathcal{S} (\cdot; \theta_s) \)~\cite{Fang2023} where the Teacher model is developed to have similar insights to the Students', aiding in efficient generalization to new data. In the first stage, the Teacher and student networks are initialized using a pre-trained source localizer, $\mathcal{M}(\cdot; \theta)$ with a regression loss, minimizing the average absolute differences from the predicted and ground-truth parameters is shown in the Equ.~\eqref{eq:regressor_loss}.

In this approach, the teacher network receives the unlabeled target data \( Z_t \), while the student network receives an augmented version by using an additive Gaussian noise \( \mathcal{N}(0, \sigma^2) \) with a pre-determined variance $\sigma^2$\cite{huang2024knowledge}. teacher network \( \mathcal{T}(\cdot; \theta_t) \) generates pseudo labels, $(X_t^{\prime},Y_t^{\prime} )$, for each sample received from $Z_t$. These newly derived pseudo labels are considered as the ground truth for Student model \( \mathcal{S}(\cdot; \theta_s) \) which would be used while proceeding with predictions, \( (\hat{X}_t, \hat{Y}_t)\). During the training process, the teacher network's parameters are updated by applying an Exponential Moving Average (EMA) process based on the Students', with a pre-determined decay rate $\alpha$. As the student network is trained to imitate the Teacher's performance and replicate its generated pseudo labels, by applying the loss function \eqref{L_KD}, any data degradation would be minimized, resulting in a stable and aligned network.
\begin{equation}
\begin{aligned}
\label{L_KD}
\mathcal{L}(\theta_s) &= \mathbb{E} \left[ |X_t' - \hat{X}_t| + |Y_t' - \hat{Y}_t| \right] \\
&\approx \frac{1}{B} \sum_{b=1}^{B} \left[ \big|x_t^{\prime (b)} - \hat{x}_t^{(b)}\big| + \big|y_t^{\prime (b)} - \hat{y}_t^{(b)}\big| \right].
\end{aligned}
\end{equation}

\begin{algorithm}
\caption{Confidence-based source-free UDA with self-supervised knowledge distillation} \label{alg:model_alg1}
\begin{algorithmic}[1]
\scriptsize
\STATE \textbf{Given:} A localizer with parameters $\mathcal{M}(\cdot; \theta)$, (unlabeled) target data $z_t$, and decay rate $\alpha$
\STATE Initialize parameters of Teacher $\mathcal{T}(\cdot; \theta_t)$ \& Student $\mathcal{S}(\cdot; \theta_s)$ with $\mathcal{M}(\cdot; \theta)$
\FOR{number of adaptation steps}
    \STATE Determine confident scores, correct labels (Algorithm~\ref{alg:model_alg2})
    \STATE Get a batch of target data $\{z_t^{(b)}\}_{b=1}^B$ with associated Pseudo Labels $\{(x_t', y_t')^{(b)}\}_{b=1}^B$.
    \STATE Do data augmentation at the Student's input 
    
    $\{z_t^{(b)} + \epsilon^{(b)}\}$ where $\epsilon^{(b)} \sim \mathcal{N}(0, \sigma)$ 

    \STATE Update Student by $\nabla_{\theta_s}[\mathcal{L}{(\theta_s)}]$ (see equation \ref{L_KD})

    \STATE Update Teacher using EMA: $\theta_t \leftarrow \alpha \cdot \theta_s + (1 - \alpha) \cdot \theta_t$ 
\ENDFOR
\end{algorithmic}
\end{algorithm}

\begin{figure}[http] 
  \centering
  \includegraphics[width=0.85\linewidth]{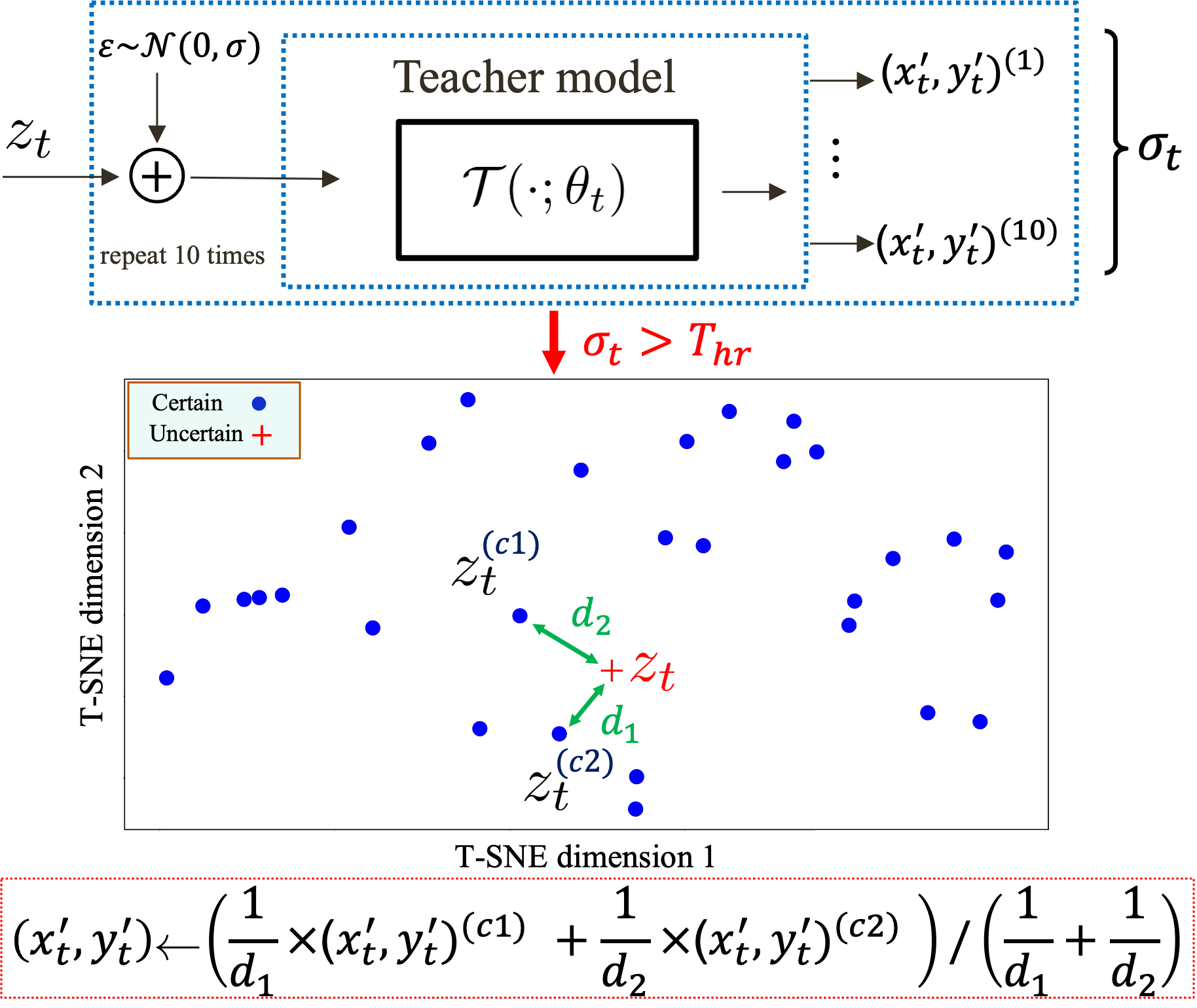} 
  \caption{\textit{MTLoc with confidence: $z_t$ together with the generated pseudo labels $(x_t',y_t')$ go through a correction process in which the uncertain labels would be identified based on a threshold $(T_{hr})$ and modified using $k$-NN (with $k=2$). The t-SNE plot shown above highlights an uncertain sample $z_t$ being corrected using two nearest certain ones. The newly modified pseudo label $(x_t',y_t')$ is considered as ground truth for our Student model to generate predictions and update the Teacher model with EMA while applying distillation loss.}}
\label{fig:confidence}
\end{figure}

To enhance the performance of our MTLoc model, we developed an enhanced approach, built on a confidence-based methodology, illustrated as Fig.~\ref{fig:confidence}, called MTLoc with confidence, in which every unconfident sample $z_t$ and its generated pseudo labels undergo a correctional process, modifying the uncertain predictions. This aids the Student model in being trained on more confident labels. In our correction module, \( \mathcal{T}(\cdot; \theta_t) \) generates pseudo labels $(x_t',y_t')$ using a noisy version of the unlabeled target data $z_t$. By repeating this process, the standard deviation $\sigma$ of the measurements is determined. Each sample is tested 10 times, and these repeated trials provide the basis for calculating the standard deviations for both $x$ and $y$. 

These uncertain variables would be corrected by using a $k$NN-based technique, if the standard deviation of generated labels is higher than an assigned Threshold ($T_{hr}$). For additional details regarding the implementation of the threshold, please refer to Appendix C. In this mechanism, $k$NN is applied in the input space, extracting $k$ nearest confident samples. Thereafter, an uncertain pseudo label $(x_t', y_t')$ would be modified by performing a weighted average of the $k$ confident neighboring labels as illustrated below:

\begin{equation}
\label{Confident_P}
(x_t', y_t') \leftarrow 
\frac{\sum_{i=1}^{K}\frac{1}{d_i} \times \left( x_t', y_t' \right)^{(c_i)} }{\sum_{i=1}^{K}\frac{1}{d_i}},
\end{equation}

\noindent with $d_i$ being the Euclidean Distance between the uncertain and the i-th closest confident samples while the  $(x_t', y_t')^{(c_i)}$ is the associated confident pseudo label.

\begin{algorithm}
\caption{Confidence scores and label generation process for SFDA approach} \label{alg:model_alg2}
\begin{algorithmic}[1]
\scriptsize
\STATE \textbf{Input:} Unlabeled target data $z_t$, Teacher Model $\mathcal{T}(\cdot; \theta_t)$
\STATE Compute $(x_t', y_t')$ for 10 runs with additive noise \(N(0, \sigma)\)
\STATE Identify uncertain pseudo labels as those with \(\sigma_t > T_{hr}\) where $\sigma_t$ is the standard deviation of the 10 runs and $T_{hr}$ is a pre-determined threshold.

\FOR{each uncertain sample \( z_t \)}
    \STATE Find the $k$-nearest confident samples by applying $k$-nearest neighbors to (input) data space.
    \STATE Correct the Pseudo Label \( (x_t', y_t') \) by using the weighted average of the labels of the confident samples (see equation \ref{Confident_P})

\ENDFOR
\end{algorithmic}
\end{algorithm}

The training along with the specifics of the pseudo label correctional process of our introduced confidence-based MTLoc model is detailed in Algorithms~\ref{alg:model_alg1} and~\ref{alg:model_alg2}.

\section{Results and Discussion}\label{sec:sectionIV}

\subsection{Experimental Setup}

\subsubsection{Environment and devices}
\begin{figure*}[htbp]
\centering
\includegraphics[width=1\textwidth]{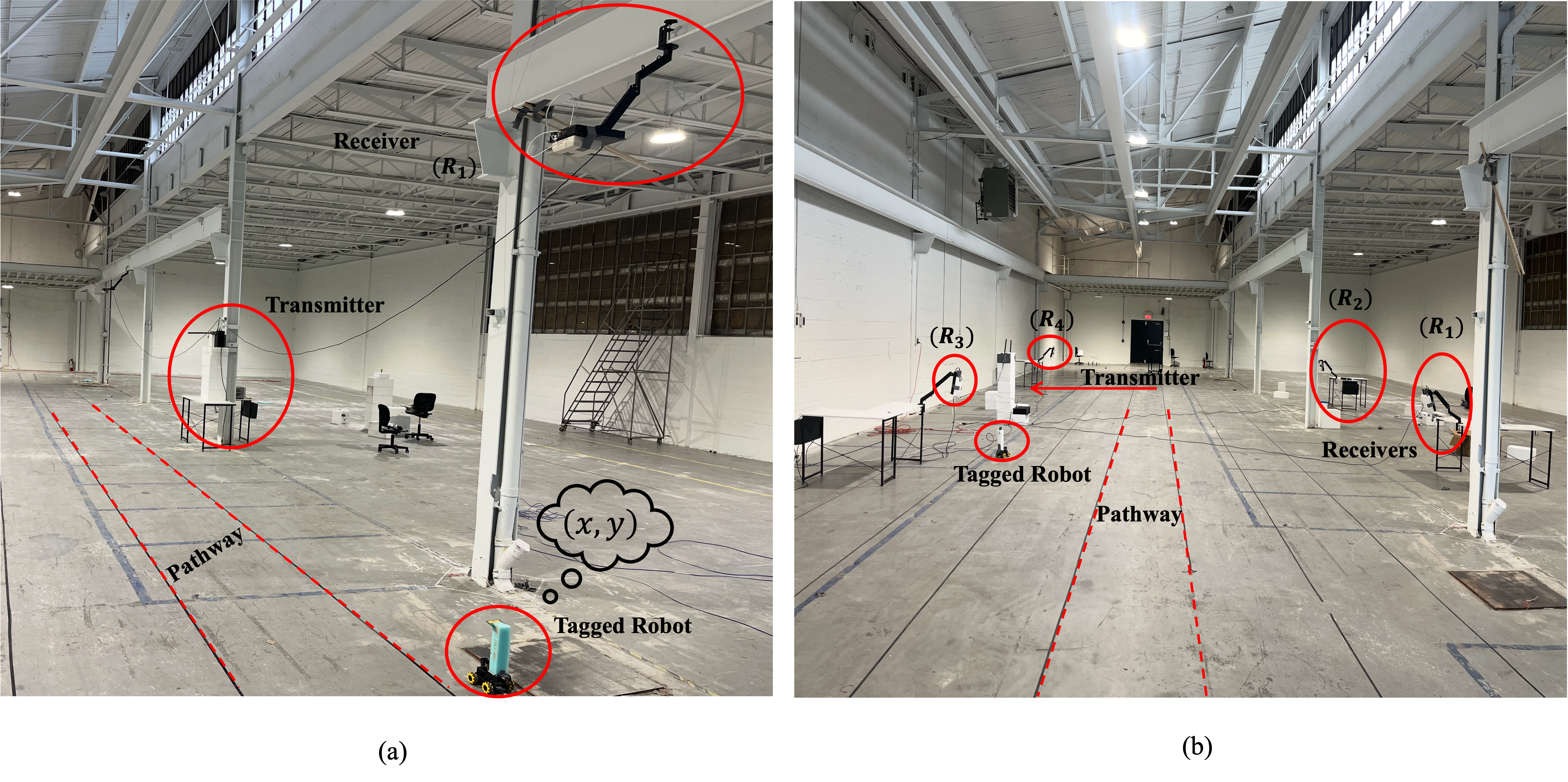}
\caption{Experimental setup of INLAN's Ceiling and Square datasets where (a) is the source domain formation with receivers positioned on the ceiling, and a transmitter activating RFID tag, located in the robot moving on pre-determined pathways and (b) the target domain's setup with receivers installed in the square position.}
\label{fig:model}
\end{figure*}

The experiments were performed in INLAN's testing facility~\cite{INLAN_data} where a transmitter and four receivers are deployed in different configurations that are specified in the datasets provided (see Fig.~\ref{fig:model}).
The transmitter is equipped with a dipole stick antenna (2 stick antennas for cross and square datasets) with 2.15 dBi gain, and each receiver is equipped with a patch antenna with 16 dBi gain. Note that for the locations of the devices, the bottom left corner of the warehouse is considered as the origin, all units are in terms of meters and the positions are given in the form $(x,y)$ where $x$ and $y$ denotes the horizontal and vertical axes, respectively. While both cross-environment and within-environment domain shifts are possible, in this work we focus on the latter by treating changes in sensor placement and layout as distinct domains. Configuration details are as follows:

\begin{itemize}
    \item Ceiling configuration: Transmitter is located at $(8.5,13)$ meter, receivers are located at the points $(8.5,5), (0.5,13.6), (8.5,21.9), (17.5,14)$. The transmitter antenna is placed parallel to the ground at a height of 1.8 m. Receivers are mounted on the ceiling at a height of 3 m with each patch antenna parallel to the ground. The tag is also parallel to the ground and elevated at a height of 0.36 m.
    \item Square configuration: Transmitter is located at $(5,9)$ meter, receivers are located at the points $(0,1.5), (0,15), (11,14), (11,2.5)$. Transmitter antennae separated by 0.4 m are placed perpendicular to the ground at a height of 1.8 m. Receivers are mounted on tables at a height of 1.4 m with each patch antenna perpendicular to the ground at a slight angle to capture the tag's movement. The tag is also perpendicular to the ground and elevated at a height of 0.4 m.
    \item Cross configuration: Transmitter is located at $(5,9)$ meter, receivers are located at the points $(0,8), (6.5,17.5), (12.5,8.8), (6.3,0)$. Transmitter antennae separated by 0.4 m are placed perpendicular to the ground at a height of 1.8 m. Receivers are mounted on tables at a height of 1.4 m with each patch antenna perpendicular to the ground at a slight angle to capture the tag's movement. The tag is also perpendicular to the ground and elevated at a height of 0.4 m.
\end{itemize}
\subsubsection{Operational specifications}
The dataset utilized in this paper was obtained using the INLAN RFID system. Unlike conventional RFID systems, which are limited to a sub-10-meter reading range, the INLAN system enables ultra-long-range reading capability over 30 meters, positioning RFID technology as a viable solution for real-time localization. This extended range is achieved by introducing harmonic RFID tags in conjunction with a distributed bi-static configuration, in contrast to the conventional mono-static setup used in RFID readers. In this setting, a transmitter operates at the 915 MHz frequency band, illuminating the RFID tag. INLAN harmonic RFID tags receive this 915 MHz signal and respond at 1.83 GHz. The response signal is captured by multiple receivers distributed across the monitoring environment. Each receiver has three heterodyne lines, which measure the signal's strength. In the presented experiments, two of the three lines measure the received power at both x- and y-polarizations. The measured data is then transferred back to the transmitter, which also acts as the system's gateway.

\subsubsection{Data collection}
Data collection is performed using a laptop running INLAN's proprietary software, connected to the transmitter via an ethernet cable. An RFID tag mounted on a car follows a pre-programmed path (comprising straight lines spaced a meter apart for the experiments), with its movement synchronized using the software to generate ground truth data. As the car moves at a constant speed of approximately 0.28 m/sec, power readings (in dBm) from each receiver (with two polarizations each) are recorded. The time interval between each reading pair of a given receiver is approximately 0.12 seconds. For the square and cross configurations, the transmitter's output is periodically switched between two antennas, and the maximum reading is recorded, effectively extending the time between consecutive readings to 0.25 seconds.

\subsection{Comparison with Baseline Models}

\begin{table*}[!htbp]

    \centering
    \caption{Performance of our MTLoc model on two target datasets (Cross \& Square), compared to source-only, DANN, ADDA, SHOT, DAFI, FiDO, and oracle. Mean Absolute Error (MAE) and Root Mean Squared Error (RMSE) values were obtained over 10 runs (excluding the source-only approach, which was pre-trained on the source dataset and evaluated on the target set).}

    \begin{adjustbox}{width=\textwidth}
    \begin{tabular}{lccccccccc}
        \toprule
        \multirow{2}{*}{\textbf{Method}} & \multirow{2}{*}{\textbf{Source-free}} & \multirow{2}{*}{\textbf{Dataset}} & \multicolumn{3}{c}{\textbf{MAE (m)} $\downarrow$} & \multicolumn{3}{c}{\textbf{RMSE (m)} $\downarrow$} \\
        \cmidrule(lr){4-6} \cmidrule(lr){7-9}
        & & & \textbf{(x)} & \textbf{(y)} & \textbf{(d)} & \textbf{(x)} & \textbf{(y)} & \textbf{(d)} \\
        \midrule \midrule
        Source-only &  & Cross  & $12.34$ & $15.50$ & $26.96$ & $15.93$ & $23.14$ & $27.29$ \\
        &  & Square & $12.37$ & $16.05$ & $16.58$ & $11.19$ & $14.72$ & $17.59$ \\
        \midrule
        DANN (inspired by FreeLoc [31]) & \xmark & Cross  & $2.22 \unc{0.01}$ & $7.59 \unc{0.01}$ & $8.18 \unc{0.01}$ & $2.81 \unc{0.01}$ & $8.95 \unc{0.01}$ & $9.30 \unc{0.01}$ \\
        & \xmark & Square & $2.32 \unc{0.01}$ & $7.64 \unc{0.01}$ & $8.26 \unc{0.01}$  & $2.91 \unc{0.01}$ & $8.99 \unc{0.01}$ & $9.37 \unc{0.01}$ \\
        \midrule
        ADDA  [73]  & \xmark & Cross  & $3.29 \unc{1.56}$ & $4.58 \unc{0.80}$ & $6.15 \unc{1.34}$ & $3.70 \unc{1.57}$ & $5.40 \unc{1.01}$ & $6.65 \unc{1.43}$ \\
        & \xmark & Square & $3.46 \unc{1.68}$ & $4.56 \unc{1.03}$ & $6.24 \unc{1.52}$ & $3.82 \unc{1.67}$ & $5.38 \unc{1.16}$ & $6.72 \unc{1.56}$ \\
        \midrule
        SHOT [74] & \xmark & Cross  & $2.96 \unc{0.29}$ & $4.14 \unc{0.08}$ & $5.55 \unc{0.29}$  & $3.48 \unc{0.33}$ & $4.91 \unc{0.16}$ & $6.02 \unc{0.31}$ \\
        & \xmark & Square & $3.05 \unc{0.29}$ & $4.14 \unc{0.03}$ & $5.59 \unc{0.23}$  & $3.58 \unc{0.35}$ & $4.87 \unc{0.05}$ & $6.05 \unc{0.23}$ \\
        \midrule
        DAFI  [30] & \xmark & Cross  & $2.59 \unc{0.64}$ & $5.48 \unc{1.57}$ & $6.46 \unc{1.70}$ & $2.99 \unc{0.69}$ & $6.50 \unc{1.80}$ & $7.17 \unc{1.86}$ \\
        & \xmark & Square & $2.83 \unc{0.75}$ & $5.28 \unc{1.32}$ & $6.47 \unc{1.07}$ & $3.24 \unc{0.78}$ & $6.28 \unc{1.61}$ & $7.17 \unc{1.30}$ \\
        \midrule
        FiDO  [29] & \xmark & Cross  & $4.51 \unc{0.77}$ & $4.54 \unc{0.61}$ & $6.91 \unc{0.95}$  & $4.96 \unc{0.82}$ & $5.59 \unc{0.93}$ & $7.48 \unc{1.17}$ \\
        & \xmark & Square & $4.29 \unc{1.56}$ & $4.90 \unc{1.40}$ & $7.10 \unc{1.20}$  & $4.64 \unc{1.59}$ & $5.83 \unc{1.53}$ & $7.65 \unc{1.22}$ \\
        \midrule
        MTLoc (Our Model) & \cmark & Cross  & $1.85 \unc{0.27}$ & $4.58 \unc{0.60}$ & $\underline{5.23} \unc{\underline{0.54}}$  & $2.18 \unc{0.28}$ & $5.43 \unc{0.82}$ & $\underline{5.86} \unc{\underline{0.74}}$ \\
        & \cmark & Square & $1.76 \unc{0.10}$ & $4.53 \unc{0.25}$ & $\underline{5.13} \unc{\underline{0.23}}$  & $2.09 \unc{0.13}$ & $5.41 \unc{0.39}$ & $\underline{5.80} \unc{\underline{0.35}}$ \\
        \midrule
        MTLoc with confidence & \cmark & Cross & $1.85 \unc{0.24}$ & $4.31 \unc{0.39}$ & $\mathbf{4.98} \unc{\mathbf{0.37}}$  & $2.18 \unc{0.26}$ & $5.11 \unc{0.53}$ & $\mathbf{5.57} \unc{\mathbf{0.50}}$ \\
        & \cmark & Square  & $1.76 \unc{0.16}$ & $4.13 \unc{0.07}$ & $\mathbf{4.75} \unc{\mathbf{0.06}}$ & $2.09 \unc{0.18}$ & $4.78 \unc{0.09}$ & $\mathbf{5.22} \unc{\mathbf{0.07}}$ \\
        \midrule
        Oracle  &  & Cross  & $0.61 \unc{0.05}$ & $1.03 \unc{0.08}$ & $1.00 \unc{0.07}$ & $0.78 \unc{0.06}$ & $1.46 \unc{0.14}$ & $1.43 \unc{0.14}$ \\
        &  & Square & $0.41 \unc{0.02}$ & $0.78 \unc{0.04}$ & $0.73 \unc{0.03}$ & $0.54 \unc{0.02}$ & $1.07 \unc{0.05}$ & $1.00 \unc{0.05}$ \\
        \bottomrule
    \end{tabular}
    \end{adjustbox}
    \label{tab:example_grouped}
\end{table*}

As previously mentioned, we analyzed our indoor localization frameworks developed based on SFDA, called MTLoc, and MTLoc with confidence, with seven baseline models: source-only, DANN, ADDA, DAFI, FiDO, SHOT and Oracle (see Table~\ref{tab:example_grouped}). We utilized INLAN's Ceiling, Cross, and Square datasets, having 12273, 1485, and 1562 samples, respectively. Under this within-environment setup, the Ceiling configuration serves as the source domain, whereas Cross and Square serve as target domains induced solely by sensor-placement changes within the same environment. The data was split into training and testing sets with an 80-20 ratio. To ensure robust evaluation, we used 5-fold cross-validation by further dividing the training set into training and validation subsets. Then, the performance of the models was assessed using the held-out test data. To identify the optimal hyperparameters, we examined the error distances corresponding to various validation losses. Through this analysis, we found that for our Cross dataset, the combination of an alpha value of 0.8, a k value of 2, and x and y threshold coefficients of 8 and 4, respectively, yielded the smallest distance error. Similarly, for the Square dataset, $x$ and $y$ threshold coefficients of 4 and 2 resulted in the minimum distance error. The implementation specifics are introduced below, and respective architectures are described in Appendix A.

\noindent \textbf{Localizer:} Our source model is trained using a one-dimensional CNN incorporating 9 layers, with INLAN's Ceiling dataset, employed in our MTLoc and baseline models.

\noindent \textbf{MTLoc:} This approach is based on the mean teacher technique using Student and teacher networks, initialized by our localizer, where our Teacher model is updated by applying EMA with an alpha of 0.7. 

\noindent \textbf{MTLoc with confidence:} Our enhanced approach follows MTLoc's methodology, using a confidence-based $k$-NN method ($k$=2) and an alpha of 0.8 to adjust uncertain pseudo labels to the nearest confident ones.

\noindent \textbf{DANN:} Our UDA baseline is created using a localizer initialized by our localizer and a discriminator encompassing three layers of 1D-CNN. 

\noindent \textbf{ADDA:} Adversarial learning based technique, aligning target features to the source using domain discriminator. 

\noindent \textbf{FiDO:} An autoencoder-based method that promotes domain-invariant features by training a reconstruction head on unlabeled target data.

\noindent \textbf{DAFI:} A GRL-based UDA framework with dual domain heads that minimizes target conditional entropy and enforces logit-space triplet alignment, while predicting continuous $(x,y)$ as the $q$-weighted average of class centers.

\noindent \textbf{SHOT:} An SFDA method for regression that freezes the regressor, adapts the feature extractor under an EMA teacher, and enforces prediction-statistics matching with CORAL-based feature alignment.

\noindent \textbf{Source-only:} In this pre-adaptation baseline, we tested our unlabeled target datasets (INLAN's Cross and Square sets) using our source localizer.

\noindent \textbf{Oracle:} In this approach, we fine-tuned our localizer with labeled target data.

\begin{figure}[htbp!] 
  \centering
  \includegraphics[width=1\linewidth]{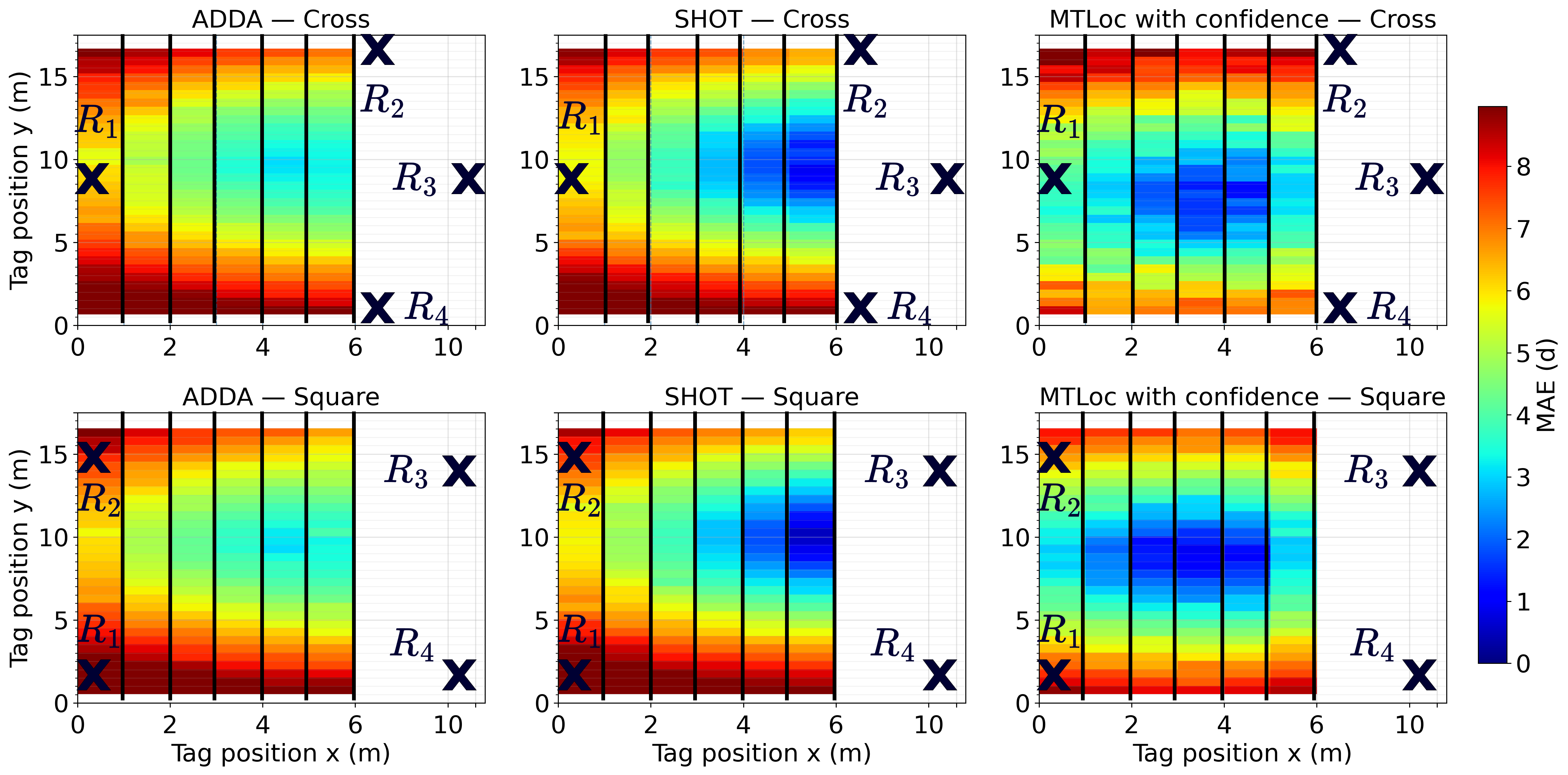} 
  \caption{Heatmaps of mean absolute distance error for MTLoc with confidence and the strongest baselines (ADDA, SHOT) on the Cross and Square target datasets. Receiver locations ($R_1-R_4$) are indicated.}
\label{fig:heatmap}
\end{figure}

As shown in Table~\ref{tab:example_grouped}, the MAE results for both the $x$ and $y$ values of the source-only model are notably high. This outcome can be due to a significant shift in the environment between training and deployment, often referred to as a domain or distributional shift. The Oracle method, which utilizes labeled data to fine-tune the source model, achieves significantly lower error, indicating the upper bound of adaptation performance. In contrast, other methods such as DANN, ADDA, DAFI, FiDO, and SHOT fall between these extremes. As the $x$ values represent the distance between pathways, we analyze the $y$ values and their combined distance $d$ to obtain a more comprehensive evaluation. Therefore, we report MAE for all methods. With the DANN baseline, MAE is reduced considerably compared to the source-only model, motivating further investigation into source-free adaptation. Our MTLoc model yields substantial improvements over all considered baselines. Compared with the baselines (ADDA, DANN, SHOT, DAFI, FiDO), MTLoc reduces MAE on average by 20.0\% and 22.5\% for Cross and Square respectively. Further improvements are observed with MTLoc with Confidence, which achieves average reductions of 23.9\% on Cross and 28.2\% on Square. Compared with individual baselines, MTLoc with Confidence achieves MAE reductions in the range of 10.3\%–27.9\% on Cross and 15.0\%–33.1\% on Square. Additionally, the standard deviation of MAE is significantly lower, with relative reductions of approximately 56.9\% on Cross and 92.6\% on Square, indicating improved robustness and consistency across runs. To better present the performance of our approach, we generate heatmaps, as shown in Fig.\ref{fig:heatmap} for our best model (MTLoc with confidence) alongside two strong baselines (ADDA and SHOT). Overall, the results show that MTLoc maintains lower localization error across both datasets (Cross \& Square), confirming its robustness and consistency in cross-domain scenarios. Notably, a confidence-based $k$-NN correction was applied during adaptation. The confidence module is active only during adaptation, ensuring there is no additional computational cost at inference. The deployed model runs identically to MTLoc during testing. While average error reductions may be modest in some cases, we consistently observe smaller standard deviations, particularly for the distance metric MAE, indicating improved stability across runs. These properties make MTLoc practical and robust for SFDA in real-world indoor localization.

\subsection{Ablation Studies}

In this section, we observe the impact of hyper-parameter values, including the EMA's alpha, $k$ of our $ k$-NN-based correctional model, and the threshold we set for our $x$ and $y$-predicted pseudo labels. For this experiment, our MTLoc with confidence model is applied on the same datasets, testing $k$ values of 1, 2, and 3, along with alpha values of 0.6, 0.7, 0.75, and 0.8. Additionally, we applied $x$ and $y$ coefficients for thresholds of 3, 4, 5, and 4, 5, and 6, respectively. The validation and test losses were observed using 5-fold cross-validation. With an alpha of 0.8 and a $k$ value of 2 obtaining optimal results on the validation set for both the Cross and Square datasets using threshold coefficients of (8,4) and (4,2), we conducted an ablation study to assess the impact of $k$ and alpha, as shown in Tables~\ref{tab:ablation_k_alpha08}\&\ref{tab:ablation_alpha_k2}.

\begin{table}[H]

\centering
\caption{\textit{Ablation Results on $\alpha$: } The test MAE values and validation distance errors are shown for the Cross and Square datasets at varying $\alpha$ values. The $x$ and $y$ threshold coefficients are set to (8,4) for the Cross and (4,2) for the Square datasets.}
\label{tab:ablation_k}
\begin{adjustbox}{width=\columnwidth}
\begin{tabular}{lccccccccc}
    \toprule
    \multirow{2}{*}{$\boldsymbol{k}$} & \multirow{2}{*}{\textbf{Dataset}} & \multicolumn{3}{c}{\textbf{MAE (m)} $\downarrow$} & \multicolumn{3}{c}{\textbf{RMSE (m)} $\downarrow$} \\
    \cmidrule(lr){3-5} \cmidrule(lr){6-8}
    & & \textbf{(x)} & \textbf{(y)} & \textbf{(d)} & \textbf{(x)} & \textbf{(y)} & \textbf{(d)} \\
    \midrule \midrule
    1 & Cross  & $1.86 \pm 0.25$ & $4.78 \pm 0.46$ & $5.42 \pm 0.42$ & $2.20 \pm 0.27$ & $5.67 \pm 0.68$ & $6.10 \pm 0.60$ \\
      & Square & $1.78 \pm 0.20$ & $4.21 \pm 0.15$ & $4.82 \pm 0.15$ & $2.09 \pm 0.22$ & $4.90 \pm 0.18$ & $5.33 \pm 0.18$ \\
    \midrule
    2 & Cross  & $1.85 \pm 0.24$ & $4.31 \pm 0.39$ & $\mathbf{4.98} \pm \mathbf{0.37}$ & $2.18 \pm 0.26$ & $5.11 \pm 0.53$ & $\mathbf{5.57} \pm \mathbf{0.50}$ \\
      & Square & $1.76 \pm 0.16$ & $4.13 \pm 0.07$ & $\mathbf{4.75} \pm \mathbf{0.06}$ & $2.09 \pm 0.18$ & $4.78 \pm 0.09$ & $\mathbf{5.22} \pm \mathbf{0.07}$ \\
    \midrule
    3 & Cross  & $1.85 \pm 0.23$ & $4.58 \pm 0.47$ & $5.24 \pm 0.40$ & $2.18 \pm 0.25$ & $5.49 \pm 0.63$ & $5.91 \pm 0.55$ \\
      & Square & $1.79 \pm 0.16$ & $4.21 \pm 0.09$ & $4.84 \pm 0.14$ & $2.14 \pm 0.17$ & $4.89 \pm 0.16$ & $5.34 \pm 0.18$ \\
    \midrule
    4 & Cross  & $2.19 \pm 0.42$ & $4.38 \pm 0.30$ & $5.22 \pm 0.37$ & $2.54 \pm 0.46$ & $5.18 \pm 0.40$ & $5.79 \pm 0.44$ \\
      & Square & $1.72 \pm 0.11$ & $4.25 \pm 0.18$ & $4.84 \pm 0.19$ & $2.05 \pm 0.12$ & $4.97 \pm 0.28$ & $5.38 \pm 0.27$ \\
    \bottomrule
\end{tabular}
\end{adjustbox}
\label{tab:ablation_k_alpha08}
\end{table}

\begin{table}[H]

\centering
\caption{\textit{Ablation Results on $\alpha$: } The test MAE values and validation distance errors are shown for the Cross and Square datasets at varying $\alpha$ values. The $x$ and $y$ threshold coefficients are set to (8,4) for the Cross and (4,2) for the Square datasets.}
\label{tab:ablation_alpha}
\begin{adjustbox}{width=\columnwidth}
\begin{tabular}{lccccccccc}
    \toprule
    \multirow{2}{*}{$\boldsymbol{\alpha}$} & \multirow{2}{*}{\textbf{Dataset}} & \multicolumn{3}{c}{\textbf{MAE (m)} $\downarrow$} & \multicolumn{3}{c}{\textbf{RMSE (m)} $\downarrow$} \\
    \cmidrule(lr){3-5} \cmidrule(lr){6-8}
    & & \textbf{(x)} & \textbf{(y)} & \textbf{(d)} & \textbf{(x)} & \textbf{(y)} & \textbf{(d)} \\
    \midrule \midrule
    0.70 & Cross  & $1.96 \pm 0.31$ & $4.62 \pm 0.47$ & $5.32 \pm 0.39$ & $2.29 \pm 0.34$ & $5.49 \pm 0.70$ & $5.97 \pm 0.59$ \\
         & Square & $1.74 \pm 0.11$ & $4.46 \pm 0.31$ & $5.05 \pm 0.34$ & $2.07 \pm 0.14$ & $5.31 \pm 0.47$ & $5.70 \pm 0.48$ \\
    \midrule
    0.75 & Cross  & $2.06 \pm 0.36$ & $4.32 \pm 0.24$ & $5.10 \pm 0.31$ & $2.40 \pm 0.38$ & $5.08 \pm 0.30$ & $5.63 \pm 0.32$ \\
         & Square & $1.79 \pm 0.18$ & $4.30 \pm 0.26$ & $4.93 \pm 0.27$ & $2.12 \pm 0.20$ & $5.06 \pm 0.35$ & $5.49 \pm 0.34$ \\
    \midrule
    0.80 & Cross  & $1.85 \pm 0.24$ & $4.31 \pm 0.39$ & $\mathbf{4.98} \pm \mathbf{0.37}$ & $2.18 \pm 0.26$ & $5.11 \pm 0.53$ & $\mathbf{5.57} \pm \mathbf{0.50}$ \\
         & Square & $1.76 \pm 0.16$ & $4.13 \pm 0.07$ & $\mathbf{4.75} \pm \mathbf{0.06}$ & $2.09 \pm 0.18$ & $4.78 \pm 0.09$ & $\mathbf{5.22} \pm \mathbf{0.07}$ \\
    \midrule
    0.90 & Cross  & $4.47 \pm 1.39$ & $5.51 \pm 1.06$ & $7.69 \pm 1.63$ & $4.91 \pm 1.39$ & $6.70 \pm 1.27$ & $8.34 \pm 1.68$ \\
         & Square & $2.93 \pm 0.57$ & $5.00 \pm 0.28$ & $6.28 \pm 0.56$ & $3.38 \pm 0.63$ & $6.15 \pm 0.36$ & $7.03 \pm 0.56$ \\
    \bottomrule
\end{tabular}
\end{adjustbox}
\label{tab:ablation_alpha_k2}
\end{table}

\subsection{Discussion}

MTLoc is the first SFDA method developed for indoor localization, addressing the limited availability of source data under privacy or resource constraints. The approach outperforms state-of-the-art baselines, including UDA methods (DANN, ADDA, FiDO, and DAFI) and SHOT, a source-free classification method adapted for regression. MTLoc introduces several promising foundations for further extension. While the MTLoc model with confidence shows robust performance, it still falls short of an oracle model having access to labeled target data. A future direction could be studying the case where a small amount of labeled target samples is available during adaptation. In this setting, semi-supervised and weakly supervised domain adaptation techniques could be applied. In many real-world scenarios, it is feasible to collect a few labeled reference tags in the target domain. By including 5 to 10 labeled samples can significantly improve the model's alignment with the target environment. These labeled points can serve as anchors to guide the model's predictions toward the correct locations and help stabilize pseudo-labels. This, in turn, reduces the domain shift between source and target distributions. By using a small amount of supervision in an originally unsupervised SFDA setting, this extension could improve model accuracy while maintaining the low computational cost that makes MTLoc practical for large-scale deployment.

\section{Conclusions}

Data privacy, storage, transfer costs, and domain shifts are critical challenges in UDA, often affecting model performance in localization tasks. To address these concerns, an SFDA model was proposed for indoor localization that preserves data privacy and reduces transfer costs. In particular, the new MTLoc model is initialized by a localizer with the architecture of a 1D-CNN. MTLoc utilizes a student-teacher architecture that improves stability through an MT mechanism, where the teacher network generates pseudo labels. An enhanced certainty-based approach called "MTLoc with confidence" selectively uses highly confident pseudo labels during training to improve performance. This technique refines the teacher network's pseudo labels by including the two nearest confident labels. In our experimental validation, the MTLoc with Confidence model achieves lower localization error than baseline methods, including DANN, ADDA, DAFI, FiDO, and the SFDA method SHOT. The confidence-based approach notably reduced errors in the $y$ dimension and improved stability in both $x$ and $y$ with small variations, resulting in a more robust and reliable model.

\section*{Acknowledgment}
This work was supported by the Natural Sciences and Engineering Research Council of Canada (NSERC), and Mitacs, with additional computational resources provided by the Digital Research Alliance of Canada.

\bibliographystyle{unsrt}
\bibliography{main}

@ARTICLE{Rao2024,
  author={X. Rao and Z. Luo and Y. Luo and Y. Yi and G. Lei and Y. Cao},
  journal={IEEE IoTJ},
  title={MFFALoc: CSI-Based Multifeatures Fusion Adaptive Device-Free Passive Indoor Fingerprinting Localization},
  year={2024},
  volume={11},
  number={8},
  doi={10.1109/JIOT.2023.3339797}
}

@ARTICLE{Ye2023,
  author={Q. Ye},
  journal={IEEE TNSE},
  title={SE-Loc: Security-Enhanced Indoor Localization With Semi-Supervised Deep Learning},
  year={2023},
  volume={10},
  number={5},
  doi={10.1109/TNSE.2022.3174674}
}

@ARTICLE{Ruan2023,
  author={Y. Ruan and L. Chen and X. Zhou and Z. Liu and X. Liu and G. Guo and R. Chen},
  journal={IEEE IoTJ},
  title={iPos-5G: Indoor Positioning via Commercial 5G NR CSI},
  year={2023},
  volume={10},
  number={10},
  doi={10.1109/JIOT.2022.3232221}
}

@ARTICLE{Li2023,
  author={Y. Li and S. Mazuelas and Y. Shen},
  journal={IEEE TWC},
  title={A Variational Learning Approach for Concurrent Distance Estimation and Environmental Identification},
  year={2023},
  volume={22},
  number={9},
  doi={10.1109/TWC.2023.3241178}
}

@ARTICLE{Wang2024,
  author={T. Wang and Y. Li and J. Liu and K. Hu and Y. Shen},
  journal={IEEE TWC},
  title={Multipath-Assisted Single-Anchor Localization via Deep Variational Learning},
  year={2024},
  doi={10.1109/TWC.2024.3359047}
}

@ARTICLE{Ayinla2024,
  author={S. L. Ayinla and A. A. Aziz and M. Drieberg},
  journal={IEEE Access},
  title={SALLoc: An Accurate Target Localization in WiFi-Enabled Indoor Environments via SAE-ALSTM},
  year={2024},
  volume={12},
  doi={10.1109/ACCESS.2024.3360228}
}

@ARTICLE{Liu2024,
  author={M. Liu and X. Liao and Z. Gao and Q. Li},
  journal={IEEE IoTJ},
  title={FT-Loc: A Fine-Grained Temporal Features-Based Fusion Network for Indoor Localization},
  year={2024},
  volume={11},
  number={3},
  doi={10.1109/JIOT.2023.3299625}
}

@ARTICLE{Junoh2024,
  author={S. A. Junoh and J.-Y. Pyun},
  journal={IEEE Access},
  title={Augmentation of Fingerprints for Indoor BLE Localization Using Conditional GANs},
  year={2024},
  volume={12},
  doi={10.1109/ACCESS.2024.3368449}
}

@ARTICLE{Li2021,
  author={Q. Li and H. Qu and Z. Liu and N. Zhou and W. Sun and S. Sigg and J. Li},
  journal={IEEE TETCI},
  title={AF-DCGAN: Amplitude Feature Deep Convolutional GAN for Fingerprint Construction in Indoor Localization Systems},
  year={2021},
  volume={5},
  number={3},
  doi={10.1109/TETCI.2019.2948058}
}

@ARTICLE{Wu2023,
  author={W. Wu},
  journal={IEEE Communications Letters},
  title={A Meta-Learning Approach for Device-Free Indoor Localization},
  year={2023},
  volume={27},
  number={3},
  doi={10.1109/LCOMM.2023.3241658}
}

@ARTICLE{Zhang2023,
  author={M. Zhang and Z. Fan and R. Shibasaki and X. Song},
  journal={IEEE IoTJ},
  title={Domain Adversarial Graph Convolutional Network Based on RSSI and Crowdsensing for Indoor Localization},
  year={2023},
  volume={10},
  number={15},
  doi={10.1109/JIOT.2023.3262740}
}

@ARTICLE{Yan2024,
  author={D. Yan and F. Shang and P. Yang and F. Han and Y. Yan and X.-Y. Li},
  journal={IEEE IoTJ},
  title={freeLoc: Wireless-Based Cross-Domain Device-Free Fingerprints Localization to Free User’s Motions},
  year={2024},
  doi={10.1109/JIOT.2024.3390748}
}

@ARTICLE{Zhang2024a,
  author={L. Zhang and S. Wu and T. Zhang and Q. Zhang},
  journal={IEEE IoTJ},
  title={RobLoc: Robust Wireless Localization With Dynamic Self-Adaptive Learning},
  year={2024},
  volume={11},
  number={10},
  doi={10.1109/JIOT.2024.3361253}
}

@ARTICLE{Prasad2024,
  author={G. Prasad and A. Pandey and S. Kumar},
  journal={IEEE TNSE},
  title={Domain Adaptation for Localization Using Combined Autoencoder and Gradient Reversal Layer in Dynamic IoT Environment},
  year={2024},
  volume={11},
  number={1},
  doi={10.1109/TNSE.2023.3304986}
}

@ARTICLE{Chen2022,
  author={X. Chen and H. Li and C. Zhou and X. Liu and D. Wu and G. Dudek},
  journal={IEEE IoTJ},
  title={Fidora: Robust WiFi-Based Indoor Localization via Unsupervised Domain Adaptation},
  year={2022},
  volume={9},
  number={12},
  doi={10.1109/JIOT.2022.3163391}
}

@ARTICLE{Fang2023,
  author={Y. Fang and P.-T. Yap and W. Lin and H. Zhu and M. Liu},
  journal={Neural Networks},
  title={Source-free unsupervised domain adaptation: A survey},
  year={2024},
  volume={164},
  doi={10.1016/j.neunet.2023.06.001}
}

@INPROCEEDINGS{chen2020fido,
  author={X. Chen and H. Li and C. Zhou and X. Liu and D. Wu and G. Dudek},
  title={{FiDo}: Ubiquitous fine-grained WiFi-based localization for unlabelled users via domain adaptation},
  publisher = {Association for Computing Machinery},
  year={2020}
}

@ARTICLE{li2021dafi,
  author={H. Li and X. Chen and J. Wang and D. Wu and X. Liu},
  journal={Proceedings of the ACM on IMWUT },
  title={{DAFI}: WiFi-based device-free indoor localization via domain adaptation},
  year={2021},
  volume={5},
  number={4}
}

@ARTICLE{9200758,
  author={Y. Kim and D. Cho and S. Hong},
  journal={IEEE SPL}, 
  title={Towards Privacy-Preserving Domain Adaptation}, 
  year={2020},
  volume={27},
  doi={10.1109/LSP.2020.3025112}
}

@ARTICLE{8879523,
  author={Y. Zhao and J. Xu and J. Wu and J. Hao and H. Qian},
  journal={IEEE IoTJ}, 
  title={Enhancing Camera-Based Multimodal Indoor Localization With Device-Free Movement Measurement Using WiFi}, 
  year={2020},
  volume={7},
  number={2},
  doi={10.1109/JIOT.2019.2948605}
}

@ARTICLE{10000400,
  author={H. Huang and J. Yang and X. Fang and H. Jiang and L. Xie},
  journal={IEEE IoTJ}, 
  title={VariFi: Variational Inference for Indoor Pedestrian Localization and Tracking Using IMU and WiFi RSS}, 
  year={2023},
  volume={10},
  number={10},
  doi={10.1109/JIOT.2022.3232740}
}

@ARTICLE{8918264,
  author={J. H. Seong and D. H. Seo},
  journal={IEEE IoTJ}, 
  title={Selective Unsupervised Learning-Based Wi-Fi Fingerprint System Using Autoencoder and GAN}, 
  year={2020},
  volume={7},
  number={3},
  doi={10.1109/JIOT.2019.2956986}
}

@INPROCEEDINGS{9613334,
  author={Z. Liu and X. Wang and Z. Chen and M. Zhao and S. Zhang and J. Li},
  title={3D-Fingerprint Augment based on Super-Resolution for Indoor 3D WiFi Localization}, 
  booktitle={WCSP}, 
  year={2021}
}

@ARTICLE{10190065,
  author={Z. Xu and B. Huang and B. Jia and G. Mao},
  journal={IEEE IoTJ}, 
  title={Enhancing WiFi Fingerprinting Localization Through a Co-Teaching Approach Using Crowdsourced Sequential RSS and IMU Data}, 
  year={2024},
  volume={11},
  number={2},
  doi={10.1109/JIOT.2023.3297521}
}

@INPROCEEDINGS{yang2014tagoram,
  author={L. Yang and Y. Chen and X.-Y. Li and C. Xiao and M. Li and Y. Liu},
  title={Tagoram: Real-time tracking of mobile RFID tags to high precision using COTS devices},
  booktitle={MobiCom},
  year={2014}
}

@ARTICLE{huang2023indoor,
  author={Y. Huang and S. Mazuelas and F. Ge and Y. Shen},
  journal={IEEE TMC},
  title={Indoor localization system with NLOS mitigation based on self-training},
  year={2023},
  volume={22},
  number={7}
}

@ARTICLE{9343321,
  author={Y. Yu and R. Chen and L. Chen and X. Zheng and D. Wu and W. Li and Y. Wu},
  journal={IEEE IoTJ}, 
  title={A Novel 3-D Indoor Localization Algorithm Based on BLE and Multiple Sensors}, 
  year={2021},
  volume={8},
  number={11},
  doi={10.1109/JIOT.2021.3055794}
}

@ARTICLE{10472153,
  author={L. Terças and H. Alves and C. H. M. de Lima and M. Juntti},
  journal={IEEE IoTJ}, 
  title={Bayesian-Based Indoor Factory Positioning Using AOA, TDOA, and Hybrid Measurements}, 
  year={2024},
  volume={11},
  number={12},
  doi={10.1109/JIOT.2024.3374457}
}

@ARTICLE{9903809,
  author={D.-H. Kim and A. Farhad and J.-Y. Pyun},
  journal={IEEE IoTJ}, 
  title={UWB Positioning System Based on LSTM Classification With Mitigated NLOS Effects}, 
  year={2023},
  volume={10},
  number={2},
  doi={10.1109/JIOT.2022.3209735}
}

@ARTICLE{9244574,
  author={R. Elbakly and M. Youssef},
  journal={IEEE TMC}, 
  title={Robust Low-Overhead RF-Based Localization for Realistic Environments}, 
  year={2022},
  volume={21},
  number={6},
  doi={10.1109/TMC.2020.3034620}
}

@INPROCEEDINGS{varailhon2024sourcefreedomainadaptationyolo,
  author={S. Varailhon and M. Aminbeidokhti and M. Pedersoli and E. Granger},
  title={Source-Free Domain Adaptation for YOLO Object Detection}, 
  booktitle={ECCV},
  year={2024}
}

@INPROCEEDINGS{Belal_2024_WACV,
  author={A. Belal and A. Meethal and F. P. Romero and M. Pedersoli and E. Granger},
  title={Multi-Source Domain Adaptation for Object Detection With Prototype-Based Mean Teacher},
  booktitle={WACV},
  year={2024}
}

@ARTICLE{10416167,
  author={Y. Etiabi and E. M. Amhoud},
  journal={IEEE Sensors Journal}, 
  title={Federated Distillation Based Indoor Localization for IoT Networks}, 
  year={2024},
  volume={24},
  number={7},
  doi={10.1109/JSEN.2024.3357798}
}

@ARTICLE{10269042,
  author={H. Liu and H. Xue and L. Zhao and D. Chen and Z. Peng and G. Zhang},
  journal={IEEE TVCG}, 
  title={MagLoc-AR: Magnetic-Based Localization for Visual-Free Augmented Reality in Large-Scale Indoor Environments}, 
  year={2023},
  volume={29},
  number={11},
  doi={10.1109/TVCG.2023.3321088}
}

@ARTICLE{6587041,
  author={W. Zhu and J. Cao and Y. Xu and L. Yang and J. Kong},
  journal={IEEE TPDS}, 
  title={Fault-Tolerant RFID Reader Localization Based on Passive RFID Tags}, 
  year={2014},
  volume={25},
  number={8},
  doi={10.1109/TPDS.2013.217}
}

@INPROCEEDINGS{5162266,
  author={C.-S. Cheng and H. H. Chang and Y.-T. Chen and T.-H. Lin and P.-C. Chen and C.-M. Huang and H. S. Yuan and W. C. Chu},
  title={Accurate Location Tracking Based on Active RFID for Health and Safety Monitoring}, 
  booktitle={ICBBE}, 
  year={2009}
}

@INPROCEEDINGS{5304900,
  author={W.-H. Chen and H. H. Chang and T.-H. Lin and P.-C. Chen and L.-K. Chen and S. J. Hwang and D. H. Yen and H. S. Yuan and W. C. Chu},
  title={Dynamic Indoor Localization Based on Active RFID for Healthcare Applications: A Shape Constraint Approach}, 
  booktitle={BMEI}, 
  year={2009}
}

@ARTICLE{huang2024knowledge,
  author={T. Huang and Y. Zhang and M. Zheng and S. You and F. Wang and C. Qian and C. Xu},
  journal={NeurIPS},
  title={Knowledge diffusion for distillation},
  year={2024},
  volume={36}
}

@INPROCEEDINGS{10706359,
  author={X. Weng and K.V. Ling and H. Liu and K. Cao},
  title={Towards End-to-End GPS Localization with Neural Pseudorange Correction}, 
  booktitle={FUSION}, 
  year={2024}
}

@ARTICLE{10452835,
  author={J. Li and Z. Yu and Z. Du and L. Zhu and H. T. Shen},
  journal={IEEE TPAMI}, 
  title={A Comprehensive Survey on Source-Free Domain Adaptation}, 
  year={2024},
  volume={46},
  number={8},
  doi={10.1109/TPAMI.2024.3370978}
}

@ARTICLE{10078842,
  author={J. Pei and Z. Jiang and A. Men and L. Chen and Y. Liu and Q. Chen},
  journal={IEEE TIP}, 
  title={Uncertainty-Induced Transferability Representation for Source-Free Unsupervised Domain Adaptation}, 
  year={2023},
  volume={32},
  doi={10.1109/TIP.2023.3258753}
}

@BOOK{bishop2006pattern,
  author={C. M. Bishop and N. M. Nasrabadi},
  title={Pattern recognition and machine learning},
  publisher={Springer},
  year={2006},
  volume={4},
  number={4}
}

@ARTICLE{Xu2024,
  author={Z. Xu and B. Huang and B. Jia and G. Mao},
  journal={IEEE IoTJ},
  title={Enhancing WiFi Fingerprinting Localization Through a Co-Teaching Approach Using Crowdsourced Sequential RSS and IMU Data},
  year={2024},
  volume={11},
  number={2},
  doi={10.1109/JIOT.2023.3297521}
}

@ARTICLE{10745164,
  author={L. Zhang and S. Wu and T. Zhang and Q. Zhang},
  journal={IEEE TII}, 
  title={Automatic Radio Map Adaptation for Robust Indoor Localization With Dynamic Adversarial Learning}, 
  year={2024},
  doi={10.1109/TII.2024.3485769}
}

@INPROCEEDINGS{Yu2024,
  author={Y. Yu and S. Shin and S. Back and M. Ko and S. Noh and K. Lee},
  title={Domain-Specific Block Selection and Paired-View Pseudo-Labeling for Online Test-Time Adaptation},
  booktitle={CVPR},
  year={2024}
}

@INPROCEEDINGS{Ganin2015DANN,
  author={Y. Ganin and V. Lempitsky},
  title={Unsupervised Domain Adaptation by Backpropagation},
  booktitle={Proc. of the 32nd Int. Conf. Mach. Learn.},
  year={2015},
  
}

@INPROCEEDINGS{Long2015,
  author={M. Long and Y. Cao and J. Wang and M. Jordan},
  title={Learning transferable features with deep adaptation networks},
  booktitle={ICML},
  year={2015},
  
}

@INPROCEEDINGS{Cui2020,
  author={S. Cui and S. Wang and J. Zhuo and L. Li and Q. Huang and Q. Tian},
  title={Towards discriminability and diversity: Batch nuclear-norm maximization under label insufficient situations},
  booktitle={CVPR},
  year={2020},
  
}

@ARTICLE{Shui2023,
  author={C. Shui and et al.},
  title={Towards more general loss and setting in unsupervised domain adaptation},
  journal={IEEE TKDE},
  year={2023},
  volume={35},
  number={10},
  pages={10140--10150},
  month={Oct.}
}

@INPROCEEDINGS{Long2018,
  author={M. Long and Z. Cao and J. Wang and M. I. Jordan},
  title={Conditional adversarial domain adaptation},
  booktitle={NeurIPS},
  year={2018},

}

@INPROCEEDINGS{Tzeng2017,
  author={E. Tzeng and J. Hoffman and K. Saenko and T. Darrell},
  title={Adversarial discriminative domain adaptation},
  booktitle={CVPR},
  year={2017},
  
}

@ARTICLE{Li2022,
  author={J. Li and M. Jing and H. Su and K. Lu and L. Zhu and H. T. Shen},
  title={Faster domain adaptation networks},
  journal={IEEE TKDE},
  year={2022},
  volume={34},
  number={12},
  pages={5770--5783},
  month={Dec.}
}

@INPROCEEDINGS{Lian2019,
  author={Q. Lian and F. Lv and L. Duan and B. Gong},
  title={Constructing self-motivated pyramid curriculums for cross-domain semantic segmentation: A non-adversarial approach},
  booktitle={CVPR},
  year={2019},
 
}

@INPROCEEDINGS{Zou2018,
  author={Y. Zou and Z. Yu and B. Kumar and J. Wang},
  title={Unsupervised domain adaptation for semantic segmentation via class-balanced self-training},
  booktitle={ECCV},
  year={2018},
 
}

@ARTICLE{Li2022Divergence,
  author={J. Li and Z. Du and L. Zhu and Z. Ding and K. Lu and H. T. Shen},
  title={Divergence-agnostic unsupervised domain adaptation by adversarial attacks},
  journal={IEEE TPAMI},
  year={2022},
  volume={44},
  number={11},
  pages={8196--8211},
  month={Nov.}
}

@ARTICLE{Li2024,
  author={J. Li and Z. Yu and Z. Du and L. Zhu and H. T. Shen},
  title={A Comprehensive Survey on Source-Free Domain Adaptation},
  journal={IEEE TPAMI},
  year={2024},
  volume={46},
  number={8},
  pages={5743--5762},
  doi={10.1109/TPAMI.2024.3370978}
}

@ARTICLE{10756560,
  author={Y. Li and B. Yu and L. Huang},
  title={An Indoor UWB Localization Method Based on Adaptive Channel Bias Estimation},
  journal={IEEE Sensors Journal},
  year={2025},
  volume={25},
  number={1},
  pages={1339--1349},
  doi={10.1109/JSEN.2024.3493069}
}

@ARTICLE{7842591,
  author={Z. Gao and Y. Ma and K. Liu and X. Miao and Y. Zhao},
  title={An Indoor Multi-Tag Cooperative Localization Algorithm Based on NMDS for RFID},
  journal={IEEE Sensors Journal},
  year={2017},
  volume={17},
  number={7},
  pages={2120--2128},
  doi={10.1109/JSEN.2017.2664338}
}

@ARTICLE{9075151,
  author={T. M. T. Dinh and N. S. Duong and K. Sandrasegaran},
  title={Smartphone-Based Indoor Positioning Using BLE iBeacon and Reliable Lightweight Fingerprint Map},
  journal={IEEE Sensors Journal},
  year={2020},
  volume={20},
  number={17},
  pages={10283--10294},
  doi={10.1109/JSEN.2020.2989411}
}

@ARTICLE{10705937,
  author={A. K. Panja and S. Biswas and S. Neogy and C. Chowdhury},
  title={Dimensionality Reduction Through Multiple Convolutional Channels for RSS-Based Indoor Localization},
  journal={IEEE Sensors Journal},
  year={2024},
  volume={24},
  number={22},
  pages={37482--37491},
  doi={10.1109/JSEN.2024.3470549}
}

@ARTICLE{10006709,
  author={H. Yang and Y. Wang and C. K. Seow and M. Sun and M. Si and L. Huang},
  title={UWB Sensor-Based Indoor LOS/NLOS Localization With Support Vector Machine Learning},
  journal={IEEE Sensors Journal},
  year={2023},
  volume={23},
  number={3},
  pages={2988--3004},
  doi={10.1109/JSEN.2022.3232479}
}

@ARTICLE{10838284,
  author={X. Liu and R. Wu and H. Zhang and Z. Chen and Y. Liu and T. Qiu},
  title={Graph Temporal Convolutional Network-Based WiFi Indoor Localization Using Fine-Grained CSI Fingerprint},
  journal={IEEE Sensors Journal},
  year={2025},
  volume={25},
  number={5},
  pages={9019--9033},
  doi={10.1109/JSEN.2025.3525624}
}

@ARTICLE{10260273,
  author={C. Huang and Z. Tian and W. He and K. Liu and Z. Li},
  title={Spotlight: A 3-D Indoor Localization System in Wireless Sensor Networks Based on Orientation and RSSI Measurements},
  journal={IEEE Sensors Journal},
  year={2023},
  volume={23},
  number={21},
  pages={26662--26676},
  doi={10.1109/JSEN.2023.3315790}
}

@ARTICLE{9676583,
  author={H. Ye and B. Yang and Z. Long and C. Dai},
  title={A Method of Indoor Positioning by Signal Fitting and PDDA Algorithm Using BLE AOA Device},
  journal={IEEE Sensors Journal},
  year={2022},
  volume={22},
  number={8},
  pages={7877--7887},
  doi={10.1109/JSEN.2022.3141739}
}

@ARTICLE{9555623,
  author={E. Fernando and O. De Silva and G. K. I. Mann and R. Gosine},
  title={Toward Developing an Indoor Localization System for MAVs Using Two or Three RF Range Anchors: An Observability Based Approach},
  journal={IEEE Sensors Journal},
  year={2022},
  volume={22},
  number={6},
  pages={5173--5187},
  doi={10.1109/JSEN.2021.3116930}
}

@ARTICLE{9082193,
  author={Z. Chen and M. I. AlHajri and M. Wu and N. T. Ali and R. M. Shubair},
  title={A Novel Real-Time Deep Learning Approach for Indoor Localization Based on RF Environment Identification},
  journal={IEEE Sensors Letters},
  year={2020},
  volume={4},
  number={6},
  pages={1--4},
  doi={10.1109/LSENS.2020.2991145}
}

@ARTICLE{9541367,
  author={J. Yan and L. Wan and W. Wei and X. Wu and W. P. Zhu and D. P. K. Lun},
  title={Device-Free Activity Detection and Wireless Localization Based on CNN Using Channel State Information Measurement},
  journal={IEEE Sensors Journal},
  year={2021},
  volume={21},
  number={21},
  pages={24482--24494},
  doi={10.1109/JSEN.2021.3114206}
}

@ARTICLE{8733822,
  author={Z. Chen and H. Zou and J. Yang and H. Jiang and L. Xie},
  title={WiFi Fingerprinting Indoor Localization Using Local Feature-Based Deep LSTM},
  journal={IEEE Systems Journal},
  year={2020},
  volume={14},
  number={2},
  pages={3001--3010},
  doi={10.1109/JSYST.2019.2918678}
}

@misc{INLAN_data,
  author={N. Mehregan and B. Bozkurt and E. Granger and M. Hajikhani and M. Shateri},
  journal={IEEE DataPort},
  title={INLAN-Indoor Localization},
  year={2025},
  howpublished = {IEEE DataPort},
  volume={10},
  number={5},
  notedoi={10.21227/w946-7z381109/TNSE.2022.3174674}
}

@ARTICLE{wang2025domain,
  author={C. Wang and Z. Wang and Q. Liu and H. Dong and W. Liu and X. Liu},
  title={A Comprehensive Survey on Domain Adaptation for Intelligent Fault Diagnosis},
  journal={Knowledge-Based Systems},
  year={2025},
  volume={327},
  pages={114109},
  doi={10.1016/j.knosys.2025.114109}
}

@ARTICLE{113645,
  author={Y. Cao and B. Luo and Y. Chen and L. Xu and C. Ding},
  title={Confidence-Aware Mean Teacher for semi-supervised metallographic image semantic segmentation},
  journal={Computational Materials Science},
  year={2025},
  volume={249},
  pages={113645},
  doi={10.1016/j.commatsci.2024.113645}
}

@article{Zhang2021IMKELM,
  author    = {Jie Zhang and Yanjiao Li and Wendong Xiao},
  title     = {Integrated Multiple Kernel Learning for Device-Free Localization in Cluttered Environments Using Spatiotemporal Information},
  journal   = {IEEE IoTJ},
  volume    = {8},
  number    = {6},
  pages     = {4749--4761},
  year      = {2021},
  doi       = {10.1109/JIOT.2020.3028574},
  publisher = {IEEE}
}

@ARTICLE{10930823,
  author={Zhang, Jie and Xue, Jianqiang and Li, Yanjiao and Cotton, Simon L.},
  journal={IEEE TMC}, 
  title={Leveraging Online Learning for Domain-Adaptation in Wi-Fi-Based Device-Free Localization}, 
  year={2025},
  volume={24},
  number={8},
  pages={7773-7787},
  doi={10.1109/TMC.2025.3552538}}

@INPROCEEDINGS{10597720,
  author={He, Tianlang and Xia, Zhiqiu and Chen, Jierun and Li, Haoliang and Chan, S.-H. Gary},
  booktitle={2024 IEEE 40th ICDE}, 
  title={Target-agnostic Source-free Domain Adaptation for Regression Tasks}, 
  year={2024}}

@inproceedings{Tzeng2017ADDA,
  title     = {Adversarial Discriminative Domain Adaptation},
  author    = {Tzeng, Eric and Hoffman, Judy and Saenko, Kate and Darrell, Trevor},
  booktitle = {CVPR},
  year      = {2017},
  month     = {July},
  pages     = {7167--7176},
  doi       = {10.1109/CVPR.2017.316}
}

@InProceedings{pmlr-v119-liang20a,
  title     = {Do We Really Need to Access the Source Data? {S}ource Hypothesis Transfer for Unsupervised Domain Adaptation},
  author    = {Liang, Jian and Hu, Dapeng and Feng, Jiashi},
  booktitle = {ICML},
  year      = {2020},

}

\section{Appendix A: Models Implementation}

In this section, implementation details of the MTLoc and MTLoc with confidence, along with the three baselines, Adversarial UDA, source-only, and oracle models, are provided. In this study, a localizer, pre-trained on the INLAN's Ceiling dataset, was designed and used in our source-only, UDA baseline and proposed SFDA approaches. An Adam optimizer with a learning rate of 0.001 was used for the optimization.

\textbf{Source-only:} In our source-only model, we used a localizer, having a feature extractor and a regressor.

\textit{Feature Extractor:}

\begin{flushleft}
\hangindent=1.5em
\hangafter=1
- Conv1D (filters = 64, kernel size = 2, activation = ‘ReLU’)

\hangindent=1.5em
\hangafter=1
- Dropout (rate = 0.2)

\hangindent=1.5em
\hangafter=1
- Conv1D (filters = 128, kernel size = 2, activation = ‘ReLU’)

\hangindent=1.5em
\hangafter=1
- Dropout (rate = 0.2)

\hangindent=1.5em
\hangafter=1
- Flatten
\end{flushleft}

\textit{Regressor:}

\begin{flushleft}
\hangindent=1.5em
\hangafter=1
- Dense (units = 128, activation = ‘ReLU’)

\hangindent=1.5em
\hangafter=1
- Dropout (rate = 0.2)

\hangindent=1.5em
\hangafter=1
- Dense (units = 64, activation = ‘ReLU’)

\hangindent=1.5em
\hangafter=1
- Dense (units = 2)
\end{flushleft}

\textbf{Adversarial UDA:} In our Adversarial UDA baseline, we used a localizer having a feature extractor and regressor with the same structure as our source-only model, together with a discriminator.

\textit{Discriminator:}

\begin{flushleft}
\hangindent=1.5em
\hangafter=1
- Dense (units = 128, activation = ‘ReLU’)

\hangindent=1.5em
\hangafter=1
- Dense (units = 64, activation = ‘ReLU’)

\hangindent=1.5em
\hangafter=1
- Dense (units = 1, activation = ‘Sigmoid’)
\end{flushleft}

\textbf{MTloc:} In our MTLoc model, our teacher and student networks have the same structure as our localizer. As the teacher network receives INLAN's Cross and Squares datasets, the Student model obtains noisy versions of them as unlabeled target data with a variance of 0.1. By applying EMA with an alpha of 0.7, the teacher networks' parameters aim to be close to the student's parameters. In our confidence-based model, following the same methodology, a threshold was defined in which the average of the predicted two-dimensional positional coordinates $(x,y)$ is added to 5 and 3 times their respective standard deviation, correcting uncertain pseudo labels by applying $k$NN ($k$=2).   

\textbf{Oracle:} In the Oracle model, we finetune our localizer, previously presented, on labeled target data (INLAN's Cross and Square datasets). 

\section{ Appendix B: Threshold Ablation Study}

In this section, threshold hyperparameters of our MTLoc with confidence model are experimentally analyzed on both the Cross and Square datasets, to better understand the behavior of the proposed approach. Table \ref{tab:ablation_k2_alpha08_full} presents the average test results of each experiment across 10 runs, along with the corresponding standard deviations obtained via 5-fold cross-validation. To define a confidence threshold for the positional coordinates, we independently scale the standard deviations of the \(x\)- and \(y\)-coordinate predictions with respect to their means and respective uncertainty distributions. This scaling factor was regarded as a tunable hyperparameter, with values of 4, 5, 6, and 8 evaluated for the \(x\) positional coordinate, and 2, 3, and 4 for the \(y\) positional coordinate. In our approach, using the cross and square datasets, we derived matrices of dimensions \(1485 \times 1\) and \(1562 \times 1\) for the \(x\) and \(y\) coordinates, respectively. Within the correction module, each training sample was processed 10 times, with random noise added in each iteration. This process resulted in matrices of dimensions \(1485 \times 10 \times 2\) and \(1562 \times 10 \times 2\), where the last dimension represents the \(x\) and \(y\) coordinates. After obtaining the outputs, we calculated the standard deviation across the 10 noise-augmented iterations for each sample, yielding \(\sigma_x\) and \(\sigma_y\) for the \(x\) and \(y\) coordinates, respectively. Our threshold is defined as:

\begin{equation}
\textit{${T_{hr}}$} = \mu + c \cdot \sigma,
\end{equation}

\noindent where \(\mu\) is the average of all the \(\sigma_x\) or \(\sigma_y\), \(\sigma\) is the standard deviation of all \(\sigma_x\) or \(\sigma_y\), and \(c\) is the scaling coefficient (\(c_x\) or \(c_y\)). This threshold identifies samples with high uncertainty, enabling the correction module to modify these pseudo labels specifically. This approach ensures focus on improving the most unconfident predictions, thereby improving its overall robustness. Table IV shows an ablation study on threshold coefficients when $k$ and alpha are 2 and 0.8, respectively.

\renewcommand{\thetable}{A-\arabic{table}}
\begin{table}[H]
\centering
\caption{\textit{Performance of MTLoc model on two target datasets (Cross \& Square) over threshold coefficients \( c_x \) and \( c_y \), for \( k = 2 \) and \( \alpha = 0.8 \).}}
\label{tab:ablation_cxcy}
\begin{adjustbox}{width=\columnwidth}
\begin{tabular}{lccccccccc}
    \toprule
    \multirow{2}{*}{$(c_x, c_y)$} & \multirow{2}{*}{\textbf{Dataset}} & \multicolumn{3}{c}{\textbf{MAE (m)} $\downarrow$} & \multicolumn{3}{c}{\textbf{RMSE (m)} $\downarrow$} \\
    \cmidrule(lr){3-5} \cmidrule(lr){6-8}
    & & \textbf{(x)} & \textbf{(y)} & \textbf{(d)} & \textbf{(x)} & \textbf{(y)} & \textbf{(d)} \\
    \midrule \midrule
(4,2) & Cross  & $1.78 \pm 0.13$ & $4.75 \pm 0.58$ & $5.36 \pm 0.51$ & $2.13 \pm 0.13$ & $5.65 \pm 0.80$ & $6.05 \pm 0.72$ \\
      & Square & $1.76 \pm 0.16$ & $4.13 \pm 0.07$ & $\mathbf{4.75} \pm \mathbf{0.06}$ & $2.09 \pm 0.18$ & $4.78 \pm 0.09$ & $\mathbf{5.22} \pm \mathbf{0.07}$ \\
\midrule
(4,3) & Cross  & $1.96 \pm 0.31$ & $4.41 \pm 0.37$ & $5.23 \pm 0.32$ & $2.30 \pm 0.34$ & $5.22 \pm 0.52$ & $5.72 \pm 0.49$ \\
      & Square & $1.78 \pm 0.17$ & $4.26 \pm 0.17$ & $4.92 \pm 0.22$ & $2.11 \pm 0.18$ & $5.01 \pm 0.27$ & $5.43 \pm 0.28$ \\
\midrule
(4,4) & Cross  & $1.90 \pm 0.38$ & $4.56 \pm 0.35$ & $5.24 \pm 0.34$ & $2.18 \pm 0.40$ & $5.42 \pm 0.50$ & $5.88 \pm 0.45$ \\
      & Square & $1.85 \pm 0.30$ & $4.19 \pm 0.10$ & $4.86 \pm 0.22$ & $2.13 \pm 0.37$ & $4.93 \pm 0.12$ & $5.35 \pm 0.25$ \\
\midrule
(5,2) & Cross  & $2.03 \pm 0.35$ & $4.57 \pm 0.56$ & $5.32 \pm 0.50$ & $2.38 \pm 0.40$ & $5.45 \pm 0.75$ & $5.97 \pm 0.67$ \\
      & Square & $1.77 \pm 0.12$ & $4.34 \pm 0.25$ & $4.95 \pm 0.26$ & $2.10 \pm 0.15$ & $5.10 \pm 0.40$ & $5.52 \pm 0.38$ \\
\midrule
(5,3) & Cross  & $2.05 \pm 0.33$ & $4.61 \pm 0.75$ & $5.37 \pm 0.60$ & $2.41 \pm 0.38$ & $5.43 \pm 1.04$ & $5.97 \pm 0.87$ \\
      & Square & $1.86 \pm 0.18$ & $4.26 \pm 0.20$ & $4.94 \pm 0.17$ & $2.21 \pm 0.20$ & $5.04 \pm 0.30$ & $5.48 \pm 0.25$ \\
\midrule
(5,4) & Cross  & $1.81 \pm 0.19$ & $4.68 \pm 0.55$ & $5.36 \pm 0.47$ & $2.14 \pm 0.21$ & $5.66 \pm 0.96$ & $6.07 \pm 0.85$ \\
      & Square & $1.82 \pm 0.15$ & $4.39 \pm 0.35$ & $5.00 \pm 0.33$ & $2.10 \pm 0.16$ & $5.25 \pm 0.45$ & $5.63 \pm 0.43$ \\
\midrule
(6,2) & Cross  & $2.03 \pm 0.35$ & $4.34 \pm 0.39$ & $5.10 \pm 0.34$ & $2.38 \pm 0.36$ & $5.12 \pm 0.59$ & $5.66 \pm 0.49$ \\
      & Square & $1.76 \pm 0.16$ & $4.33 \pm 0.19$ & $4.94 \pm 0.20$ & $2.09 \pm 0.17$ & $5.09 \pm 0.29$ & $5.50 \pm 0.25$ \\
\midrule
(6,3) & Cross  & $1.99 \pm 0.34$ & $4.57 \pm 0.24$ & $5.29 \pm 0.19$ & $2.32 \pm 0.36$ & $5.46 \pm 0.37$ & $5.95 \pm 0.27$ \\
      & Square & $1.71 \pm 0.08$ & $4.18 \pm 0.15$ & $4.78 \pm 0.12$ & $2.03 \pm 0.07$ & $4.89 \pm 0.20$ & $5.29 \pm 0.17$ \\
\midrule
(6,4) & Cross  & $1.75 \pm 0.10$ & $4.47 \pm 0.47$ & $5.12 \pm 0.40$ & $2.07 \pm 0.12$ & $5.36 \pm 0.70$ & $5.75 \pm 0.62$ \\
      & Square & $1.79 \pm 0.09$ & $4.50 \pm 0.25$ & $5.12 \pm 0.28$ & $2.07 \pm 0.10$ & $5.30 \pm 0.39$ & $5.70 \pm 0.37$ \\
\midrule
(8,4) & Cross  & $1.85 \pm 0.24$ & $4.31 \pm 0.39$ & $\mathbf{4.98} \pm \mathbf{0.37}$ & $2.18 \pm 0.26$ & $5.11 \pm 0.53$ & $\mathbf{5.57} \pm \mathbf{0.50}$ \\
      & Square & $1.71 \pm 0.05$ & $4.24 \pm 0.11$ & $4.83 \pm 0.11$ & $2.03 \pm 0.06$ & $4.98 \pm 0.17$ & $5.38 \pm 0.16$ \\
    \bottomrule
\end{tabular}
\end{adjustbox}
\label{tab:ablation_k2_alpha08_full}
\end{table}

\section{Appendix C: Adapting SHOT To Regression}

In this section, we discuss one of the baselines, SHOT [74], a classification-based SFDA method adapted to the regression setting, as shown in Fig.~\ref{fig:SHOT-R}. In the original setting, SHOT minimizes prediction entropy to ensure feature alignment. However, entropy is not applicable to regression tasks, as predictions are continuous positional coordinates. To address this, we maintain the original SHOT framework: a feature extractor $\mathcal{F}(\cdot;\theta_s)$ and a frozen, source-pretrained regressor $\mathcal{G}(\cdot;\theta_g)$. The model is trained to predict tag locations $y$ for unlabeled target data. During adaptation, only the feature extractor $\mathcal{F}(\cdot;\theta_s)$ is trained, while $\mathcal{G}(\cdot;\theta_g)$ remains frozen.

\renewcommand{\thefigure}{A-\arabic{figure}}
\begin{figure}[htbp!] 
  \centering
  \includegraphics[width=1\linewidth]{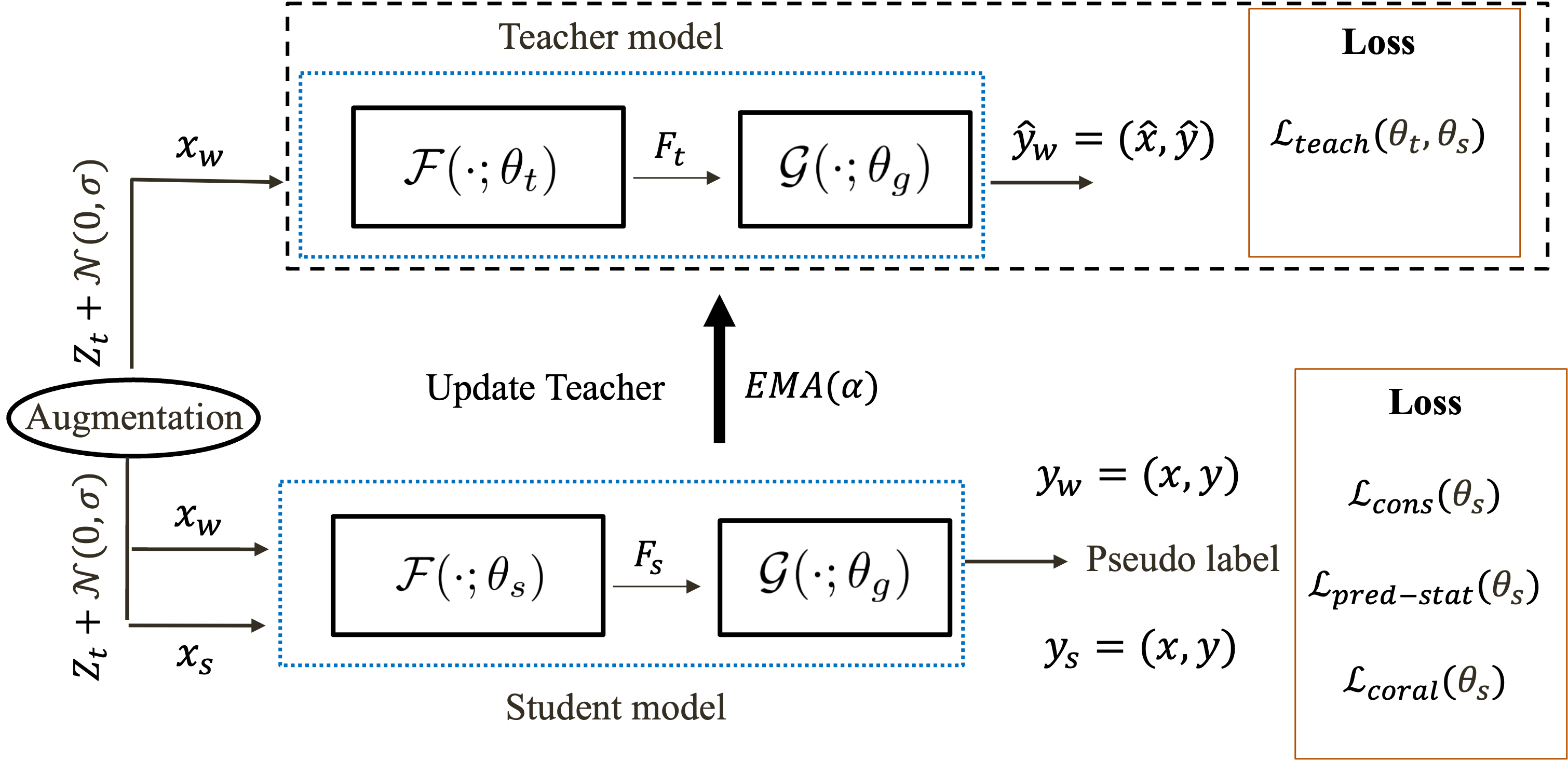} 
    \caption{\textit{Adaptation of SHOT for regression. The dashed black box indicates the teacher model, which can be optionally disregarded. Target inputs are augmented into two views: a weak view with small Gaussian noise $\mathcal{N}(0,\,0.01^2)$ and a strong view with element-wise random masking ($p=0.10$) and Gaussian noise $\mathcal{N}(0,\,0.05^2)$.}}
\label{fig:SHOT-R}
\end{figure}

For each unlabeled target sample $z_t$, two augmented views are created: a weak view $x_w = a_w(z_t)$ (small Gaussian noise, $\mathcal{N}(0,\,0.01^2)$ and a strong view $x_s = a_s(z_t)$ (element-wise random masking with probability $0.10$ having additive Gaussian noise, $\mathcal{N}(0,\,0.05^2)$, with ${y}_w$ and ${y}_s$ as their predictions. To ensure prediction consistency across views, we minimize Eq.~\ref{eq:consistency_loss}:

\begin{align}
\mathcal{L}_{\text{cons}}
&= \mathbb{E}_{x \sim \mathcal{D}_t}\!\left[ \left\| {y}_w - {y}_s \right\|_2^2 \right].
\label{eq:consistency_loss}
\end{align}

To increase the model’s stability during training, a second encoder $\mathcal{F}(\cdot;\theta_t)$ is used as a teacher model, updated via EMA: $\theta_t \leftarrow \mathrm{EMA}(\theta_s)$. In this student–teacher framework, the teacher encoder receives the weak, noisy data $x_w$ and produces target predictions. To ensure alignment between the student’s prediction on the strong view and the teacher’s outputs, we apply the teacher loss in Eq.~\ref{eq:teacher_loss}:

\begin{align}
\mathcal{L}_{\text{teach}}
&= \mathbb{E}_{x \sim \mathcal{D}_t}\!\left[ \left\| \hat{y}_s - \hat{y}_t \right\|_2^2 \right].
\label{eq:teacher_loss}
\end{align}

Furthermore, as the source-domain prediction statistics and feature covariances are computed during pre-training, statistical alignment between source and target is ensured by matching the first- and second-order moments (mean and variance) of the student’s predictions across domains via the following prediction-statistics loss (Eq.~\ref{eq:pred_stat_loss}):

\begin{align}
\mathcal{L}_{\text{pred-stat}} 
&= \left\| \mu_t - \mu_s \right\|_2^2 + \left\| \sigma_t^2 - \sigma_s^2 \right\|_2^2.
\label{eq:pred_stat_loss}
\end{align}

Additionally, to align feature-level covariance statistics across source features $\mathcal{F}_s$ and target features $\mathcal{F}_t$, we use the CORAL loss in Eq.~\ref{eq:coral_loss} to align feature distributions across domains in this source-free setting:

\begin{align}
\mathcal{L}_{\text{coral}} 
&= \left\| \operatorname{Cov}(\mathcal{F}_t) - \operatorname{Cov}(\mathcal{F}_s) \right\|_F^2,
\label{eq:coral_loss}
\end{align}

\noindent where $\|\cdot\|_F$ is the Frobenius norm. The total loss is defined as follows:

\[
\mathcal{L}_{\text{total}} =
\lambda_{\text{cons}}\mathcal{L}_{\text{cons}} +
\lambda_{\text{teach}}\mathcal{L}_{\text{teach}} +
\lambda_{\text{stat}}\mathcal{L}_{\text{pred-stat}} +
\lambda_{\text{coral}}\mathcal{L}_{\text{coral}}.
\]

\noindent This loss formulation balances consistency, statistical alignment, and teacher-guided supervision for effective SFDA. We use a learning rate of \(10^{-3}\) and an exponential moving average coefficient of \(0.995\). All loss weights are treated as hyperparameters and selected on a held-out validation split: \(\lambda_{\text{cons}}=1.0\), \(\lambda_{\text{teach}}=0.5\), \(\lambda_{\text{stat}}=0.10\), and \(\lambda_{\text{coral}}=0.02\).

\end{document}